\documentclass[onecolumn]{aastex631}

\usepackage{multirow}
\usepackage{amsmath}

\newcommand{\teff}{$T_\mathrm{eff}$}
\newcommand{\lbol}{$L_\mathrm{bol}$}
\newcommand{\Eur}{$E(u-r)$}
\newcommand{\ErHa}{$E(r-H\alpha)$}

\received{September 18, 2023}
\revised{December 18, 2023}
\accepted{January 8, 2024}

\shorttitle{Circumstellar disk accretion across the Lagoon Nebula}
\shortauthors{Venuti et al.}

\begin{document}

\title{Circumstellar disk accretion across the Lagoon Nebula: the influence of environment and stellar mass}

\author[0000-0002-4115-0318]{Laura Venuti}
\affiliation{SETI Institute, 339 Bernardo Ave., Suite 200, Mountain View, CA 94043, USA}
\affiliation{CSSM and Department of Physics, University of Adelaide, North Terrace campus, Adelaide, SA 5005, Australia}
\correspondingauthor{Laura Venuti} \email{lvenuti@seti.org}

\author[0000-0002-3656-6706]{Ann Marie Cody}
\affiliation{SETI Institute, 339 Bernardo Ave., Suite 200, Mountain View, CA 94043, USA}

\author[0000-0002-3865-9906]{Giacomo Beccari}
\affiliation{European Southern Observatory, Karl-Schwarzschild-Strasse 2, 85748 Garching bei M\"unchen, Germany}

\author[0000-0001-6381-515X]{Luisa M. Rebull}
\affiliation{Infrared Science Archive (IRSA), IPAC, 1200 E. California Blvd., California Institute of Technology, Pasadena, CA 91125, USA}

\author[0000-0002-2191-9038]{Michael J. Irwin}
\affiliation{Institute of Astronomy, University of Cambridge, Madingley Road, Cambridge CB3 0HA, UK}

\author{Apoorva Thanvantri}
\affiliation{California Institute of Technology, 1200 East California Blvd., Pasadena, CA 91125, USA}

\author{Sowmya Thanvantri}
\affiliation{University of California, Berkeley, 101 Sproul Hall, Berkeley, CA 94720, USA}

\author[0000-0002-5171-8376]{Silvia H.~P. Alencar}
\affiliation{Departamento de F\'isica -- ICEx -- UFMG, Av. Ant\^onio Carlos 6627, 30270-901 Belo Horizonte, MG, Brazil}

\author{Clara O. Leal}
\affiliation{Departamento de F\'isica -- ICEx -- UFMG, Av. Ant\^onio Carlos 6627, 30270-901 Belo Horizonte, MG, Brazil}

\author[0000-0002-3306-3484]{Geert Barentsen}
\affiliation{Bay Area Environmental Research Institute, P.O. Box 25, Moffett Field, CA 94035, USA}
\affiliation{NASA Ames Research Center, Moffett Field, CA 94035, USA}

\author[0000-0003-1192-7082]{Janet E. Drew}
\affiliation{Department of Physics \& Astronomy, University College London, Gower Street, London, WC1E 6BT, UK}

\author[0000-0002-2532-2853]{Steve B. Howell}
\affiliation{NASA Ames Research Center, Moffett Field, CA 94035, USA}

\begin{abstract}
Pre-main sequence disk accretion is pivotal in determining the final stellar properties and the early conditions for close-in planets. We aim to establish the impact of internal (stellar mass) and external (radiation field) parameters on disk evolution in the Lagoon Nebula massive star-forming region. We employ simultaneous $u,g,r,i,H\alpha$ time series photometry, archival infrared data, and high-precision {\em K2} light curves, to derive stellar, disk, and accretion properties for 1012 Lagoon Nebula members. {Of all young stars in the Lagoon Nebula}, we estimate 34\%--37\% have inner disks traceable down to $\sim 12$\,$\mu$m, while 38\%--41\% are actively accreting. {We detect disks $\sim${1.5} times more frequently around G/K/M stars than} higher-mass stars, which appear to deplete {their inner disks on shorter timescales}. We find {tentative} evidence for faster disk evolution in the central regions of the Lagoon Nebula, where the bulk of the O/B population is located. Conversely, disks appear to last longer at its outskirts, where the measured fraction of disk--bearing stars tends to exceed those of accreting and disk--free stars. The derived mass accretion rates show a non-uniform dependence on stellar mass between $\sim 0.2-5$\,$M_\odot$. In addition, the typical accretion rates appear to differ across the Lagoon Nebula extension, with values two times lower in the core region than at its periphery. Finally, we detect tentative density gradients in the accretion shocks, with lags in the appearance of brightness features as a function of wavelength that can amount to $\sim7\%-30\%$ of the rotation period.
\end{abstract}


\section{Introduction} \label{sec:intro}

During the first few million years (Myr) following the early formation stages, the evolution of young stars is governed by the interaction with their surrounding protoplanetary disks \citep[e.g.,][]{bouvier2007}. This stage plays a long-lasting role in the determination of the fundamental properties of the final star, and the dynamical {star--disk interaction can affect both} the survival and orbital configuration of close-in planets \citep{liu2017,romanova2019}. For solar-type and lower-mass young stellar objects (YSOs), the inner regions of the circumstellar disk are truncated at a distance of a few stellar radii from the central source \citep[e.g.,][]{eisner2014,gravity2020}, as a result of the pressure exerted by the intense ($\sim$kG) magnetic field at the surface of the star \citep[e.g.,][]{yang2011,johnstone2014,lavail2017}. The stellar magnetosphere therefore controls the exchange of mass and angular momentum between the star and the inner disk, via the process of magnetospheric accretion \citep{hartmann2016}. {This mechanism may also be operating in some higher--mass YSOs (among which detection of magnetic fields is less common; e.g., \citealp{alecian2013,villebrun2019})}, as suggested by the observation of rotationally--modulated line emission signatures \citep[e.g.,][]{pogodin2021,brittain2023} characteristic of magnetospheric accretion streams \citep[e.g.,][]{kurosawa2011}.  

{Determining the time dependence of accretion and} disk evolution is critical to constrain the planet formation phase and identify the disk clearing {mechanisms} \citep[e.g.,][]{alexander2006,rosotti2017}. Population--wide surveys of disk--bearing vs. accreting YSOs in open clusters as a function of age have suggested that circumstellar disks tend to disperse within 5--10~Myr \citep[e.g.,][]{hernandez2007,bell2013,ribas2014}, with the fraction of accreting stars evolving on similar or shorter timescales \citep[e.g.,][]{fedele2010,briceno2019,flaccomio2023}. However, this global picture is affected by {local differences between individual clusters}, such as the {specific mass spectrum of member stars} and the external radiation field {that can impact} the disk lifetimes \citep[e.g.,][]{guarcello2010,ribas2015,coleman2022}. In addition, a different sensitivity in accretion and disk diagnostics may bias the comparison in respective evolutionary timescales, since young stars can still be accreting even after the dust content of the inner disk has been cleared \citep{thanathibodee2022}. 

Wide--field, multi--band photometric surveys provide the most efficient approach for a uniform mapping of disk and accretion properties across large samples of YSOs. Thermal emission from circumstellar dust is responsible for the distinctive infrared (IR) flux excess in the spectral energy distribution (SED) of disk--bearing YSOs \citep[e.g.,][]{robitaille2007}, and tracing how this excess emission varies with wavelength provides a direct proxy for the disk evolutionary stage \citep[e.g.,][]{lada1987,evans2009}. Mass accretion onto the star can be studied by measuring the flux excess at short wavelengths ($<$400~nm) {emitted} by the surface shocks that develop where material is deposited from the accretion column \citep[e.g.,][]{calvet1998,orlando2010,matsakos2013}. Photometric observations in the ultraviolet (UV) provide a robust tracer of the total accretion luminosity on individual sources \citep[e.g.,][]{gullbring1998,rigliaco2011,manara2012,venuti2014}. Accretion activity is also commonly quantified by measuring the intensity of hydrogen emission lines (most prominently H$\alpha$) produced by the heated and accelerated gas in the accretion columns \citep[e.g.,][]{white2003,kurosawa2006,alcala2017}, which can be captured photometrically by adopting narrowband H$\alpha$ filters \citep[e.g.,][]{demarchi2010,barentsen2011}. {For both accretion diagnostics}, simultaneous color information in optical broadband filters is crucial for a reliable evaluation of stellar parameters, {photospheric flux, and chromospheric emission} \citep[e.g.,][]{manara2013}.

In this paper, we explore the impact of stellar mass and environmental conditions on the properties of accretion disks in YSOs by focusing on the Lagoon Nebula region. With a typical age of $<$1~Myr at its core \citep[the NGC~6530 cluster;][]{prisinzano2019}, and a rich population of a few thousand members (\citealp{feigelson2013}; Rebull et al., in preparation) and several dozen massive, O/B stars \citep{tothill2008,povich2017}, this region is an ideal target to study how these factors may disrupt the dynamics of accretion and disk evolution. 
{The Lagoon Nebula has been the object of numerous investigations across the wavelength spectrum, including recent studies on the reddening law and age spread across the region \citep{prisinzano2019}, on the spatial structure and dynamical formation history of the NGC~6530 cluster \citep{Damiani2019}, and on the instantaneous accretion properties of its solar-to-subsolar mass members \citep{Kalari2015}.}
Variability studies of Lagoon Nebula YSOs \citep[e.g.,][]{Venuti2021,ordenes2022} have shown that many cluster members are actively interacting with their circumstellar disks.
Notably, the Lagoon Nebula was one of the few young stellar fields to be encompassed by the \textit{Kepler/K2} mission \citep{howell2014}, resulting in a uniquely detailed description of the diverse variability patterns that can be found among YSOs \citep[e.g.,][]{cody2014,cody2018} and how they relate to the specific accretion and star--disk interaction dynamics in the final stages of stellar mass assembly \citep[e.g.,][]{fischer2022}.

In \citet{Venuti2021}, we presented a comprehensive study of variability behaviors and timescales for YSOs in the Lagoon Nebula by leveraging the \textit{Kepler/K2} light curves and simultaneous, multicolor time series photometry obtained from the ground in $u,g,r,i,H\alpha$ filters. That study revealed clear mass--dependent trends in the occurrence of different light curve patterns, categorized in amplitude and shape according to their degree of periodicity and symmetry as defined by \citet{cody2014}. In this paper, we take advantage of the full area coverage provided by the $u,g,r,i,H\alpha$ dataset to characterize accretion and disk properties for Lagoon Nebula YSOs across a wider region around the \textit{K2} field, and to investigate how these properties evolve as a function of stellar parameters and local conditions. {This work extends the results of previous static investigations by providing a detailed view of how star--disk characteristics within the region change across the time domain, from timescales of hours to years. We also explore the connection between accretion and disk evolution, drawing a homogeneous picture over the entire spatial extent of the Lagoon Nebula and across a sweeping range in stellar masses ($\sim 0.2-5$\,$M_\odot$).}

The paper is organized as follows. Section~\ref{sec:datasets} details the photometry acquisition procedures for our datasets. 
Section~\ref{sec:target_pop} details the selection of Lagoon Nebula members and their disk classifications.
Section~\ref{sec:stellar_properties} discusses the derivation of fundamental stellar parameters. 
Section~\ref{sec:accretion_tracers} reports on how accreting stars were identified based on UV and H$\alpha$ emission.
Section~\ref{sec:accretion_activity} explores the connection between accretion and stellar mass, the simultaneity of accretion signatures at different wavelengths, and the variability of accretion on different timescales. In Section~\ref{sec:discussion}, we discuss the implications of our findings for the pattern of disk evolution as a function of stellar mass and spatial location. Section~\ref{sec:conclusions} summarizes our conclusions and future perspectives.

\section{Observations and data processing} \label{sec:datasets}

Our work is built on multi-band light curves obtained for young stars in the Lagoon Nebula with the OmegaCAM wide-field imager \citep{kuijken2002,kuijken2011} at the VLT Survey Telescope (VST; \citealp{arnaboldi1998}). That observing campaign, described in \citet{Venuti2021}, complemented the \textit{Kepler/K2} monitoring of the NGC~6530 cluster during \textit{K2} Campaign~9. In Sect.~\ref{sec:omegacam_k2}, we summarize the technical details and light curve extraction procedure relevant to these two primary datasets, while ancillary datasets that were used to classify the evolutionary status of our target population are briefly described in Sect.~\ref{sec:auxiliary_datasets}.

\subsection{Primary datasets: VST/OmegaCAM, \textit{Kepler/K2}}\label{sec:omegacam_k2}

The VST/OmegaCAM observations were conducted between June 16 and July 10, 2016, under program 297.C-5033(A). A thorough description of that observing run and of the associated data reduction and photometry extraction procedures was provided in Sect.~2.3 of \citet{Venuti2021}. Here we summarize the key information pertaining to this dataset.

By adopting a dithered observing pattern with a sequence of two short and two long exposures for each single observing block, VST/OmegaCAM provided a five-band ($u,g,r,i,H\alpha$) map of an approximately $1.3^\circ \times 1.2^\circ$ area around (R.A.,~Dec.) $\sim$ ($271.0, -24.36$). Each star imaged with VST/OmegaCAM was observed a minimum of 6 (and generally 17) times, at a typical cadence of one visit per day over 14 non-consecutive days during the monitored period. 
Object fluxes were measured via the Cambridge Astronomy Survey Unit (CASU) pipeline\footnote{http://casu.ast.cam.ac.uk/surveys-projects/vst}, and then transformed to magnitudes and calibrated to the Sloan Digital Sky Survey (SDSS) system \citep{fukugita1996}, by taking as reference the catalog of the VST Photometric H$\alpha$ Survey of the Southern Galactic Plane and Bulge (VPHAS+; \citealp{drew2014}), which employed the same instrument and filter set.

As explained in detail in Sect.~\ref{sec:sample_selection}, a cross-correlation of our VST/OmegaCAM catalog with a literature census of Lagoon Nebula YSOs yielded a list of 1012 members for which photometric measurements are available in our dataset. For 415 of these, we could also extract high-precision time series photometry from {\em K2} Campaign 9. This subset represents an extension of the sample presented in \citet{Venuti2021}, owing to an expanded membership search (Sect.~\ref{sec:sample_selection}), and to the implementation of additional light curve extraction techniques that allowed us to enlarge the brightness range of \textit{K2} targets, as detailed below. 

The {\em K2} observations were limited to an area of $\sim$$0.26^\circ \times 0.15^\circ$ at the heart of the Lagoon Nebula (where NGC~6530 is located). For all stars in the {\em K2} sample, we created both aperture and point-spread-function (PSF) light curves. Aperture photometry for the bulk of objects was carried out as described in \citet{Venuti2021}; 
for targets that were saturated on the detector, we instead employed irregularly shaped aperture masks with the \texttt{lightkurve} Python package \citep{lightkurve}. The extent of these apertures was selected manually based on the star's flux distribution and location of saturation spikes. 

Photometry for the faintest sources on the \textit{K2} frames benefited from a PSF fitting approach, which was carried out with the Photutils Python package. An effective point spread function (ePSF) was built based on the combined flux data of eight stars from all 3284 {\em K2} Lagoon superstamp images. These eight stars were selected on criteria of brightness (Kepler magnitude range $\sim$10--15) and lack of neighbors (no sources of similar brightness within several pixels).

Stellar data, including sky positions, for stars used to create the ePSF were extracted from the Gaia archive \citep{2016A&A...595A...1G,2023A&A...674A...1G}. The same ePSF model was then used for flux fitting for each star, in each {\em K2} frame. Centroids were fixed based on each image world coordinate system (WCS) and the known star positions. 

For circular aperture and PSF-based light curves, we found that the measured flux was sensitive to object centroid position. This ``jitter" is a well known phenomenon for {\em K2} time series photometry, and it can be mitigated by estimating and removing position-correlated flux trends. We carried out this process with the \texttt{k2sc} package \citep{aigrain2016}. \texttt{k2sc} employs a Gaussian process model to estimate any underlying stellar variability in addition to flux changes caused by intrapixel sensitivity variations. Upon applying this detrending method, light curves were output with decreased systematic effects. \looseness=-1

Final {\em K2} light curves were chosen among the available versions, based on which displayed the lowest noise levels, and consequently, the clearest variability signature. Preference to PSF or photometry with the smallest apertures was given if there were companion sources present within 1--2 pixels (4--8\arcsec).

\subsection{Auxiliary datasets}\label{sec:auxiliary_datasets}

In order to classify YSOs in our sample as a function of their disk evolutionary stage, we employed near- to mid-IR photometry from three archival databases: the Two Micron All Sky Survey (2MASS; \citealp{skrutskie2006}), the {\it Spitzer/InfraRed Array Camera} (IRAC; \citealp{fazio2004}), and the Wide-field Infrared Survey Explorer (WISE; \citealp{wright2010}). Details regarding the literature sources that were merged to assemble this IR compilation are provided in \citet{Venuti2021}. Both the 2MASS and the IRAC catalogs extend over the entire $\sim$0.9 sq. deg. area over which the Lagoon Nebula population is distributed; the IRAC coverage in volume of sources is $\sim$10\% less complete than the 2MASS coverage. On the other hand, the WISE coverage is only about 15\% complete with respect to 2MASS, and the central regions of the Lagoon Nebula, where the majority of YSOs are located, are especially undersampled. 

\section{Target population}\label{sec:target_pop}

\subsection{Sample selection and cluster membership} \label{sec:sample_selection}

To identify all young stars in the VST/OmegaCAM field, we conducted a comprehensive literature search of all potential YSO members reported for the Lagoon Nebula region across the wavelength spectrum \citep[][Sect.~2.1; Rebull et al., in preparation]{Venuti2021}. {We then cross-matched the resulting list with the complete catalog of point sources extracted from the VST/OmegaCAM observations}. {This procedure led to selecting 415 confirmed or possible Lagoon Nebula members with a counterpart in the \textit{K2} mosaic, and another 597 with no \textit{K2} light curves but with VST/OmegaCAM detection} in at least the $g,r,i$ filters, to enable stellar parameter (extinction, ``A$_V$''; spectral type, ``SpT'') estimation following the approach of \citet{Venuti2021}. 

Our final YSO sample is introduced in Table~\ref{tab:sample_properties}. Overall, a complete $u,g,r,i,H\alpha$ photometric set is available for 486/1012 objects (48.0\%). This is due to the fact that the completeness level of our $u$-band catalog drops quickly on the fainter side of $r\sim 16$, which corresponds to the average $r$-band magnitude of our target list. Another 462/1012 YSOs in our sample (45.7\%) lack $u$-band photometry but have $g,r,i,H\alpha$ photometry. 
Hence, at least one accretion indicator ($u$-band excess and/or H$\alpha$ emission, as discussed in Sect.~\ref{sec:accretion_tracers}) is available for 97\% of our sample. The derived variability classes from the \textit{K2} light curves, when available, are also provided in Table~\ref{tab:sample_properties}; these are used in Sect.~\ref{sec:accretion_activity} to investigate the connection between photometric variability signatures and the dynamics and intensity of disk accretion activity.

\begin{deluxetable*}{c c c c c c c c c c c c c c c c c c c c c}
\setlength\tabcolsep{3.5pt}
\tabletypesize{\footnotesize}
\rotate
\centerwidetable
\tablecaption{List of Lagoon Nebula members included in this study, with corresponding coordinates, derived membership flag, disk classification, {accretion status,} \emph{K2} variability class (when available), VST/OmegaCAM $u,g,r,i,H\alpha$ photometry, and derived stellar and accretion parameters. \label{tab:sample_properties}}
\tablehead{\colhead{Identifier\tablenotemark{a}} & \colhead{R.A.} & \colhead{Dec.} & \colhead{Mem.} & \colhead{Disk\tablenotemark{b}} & \colhead{Acc.\tablenotemark{c}} & \colhead{\emph{K2}\tablenotemark{d}} & \colhead{$u$} & \colhead{$g$} & \colhead{$r$} & \colhead{$i$} & \colhead{$H\alpha$} & \colhead{SpT} & \colhead{A$_V$} & \colhead{\lbol} & \colhead{$M_\star$} & \colhead{$R_\star$} & \colhead{$\log{t}$} & \colhead{$\log{\dot{M}_{acc}}$} & \colhead{$\Delta\dot{M}_{+}\tablenotemark{e}$} & \colhead{$\Delta\dot{M}_{-}\tablenotemark{f}$} \\
 & [deg.] & [deg.] & & & & & [mag] & [mag] & [mag] & [mag] & [mag] & & [mag] & [$L_\odot$] & [$M_\odot$] & [$R_\odot$] & [yr] & [$M_\odot$/yr] & [dex] & [dex]
 } 
\startdata
224290871 & 270.9510 & -24.5109 & ? & y* & {0} & QPS & 18.48 & 16.54 & 15.17 & 14.49 & 14.93 & K8.3 & 0.04 & 2.35 & 0.53 & 3.23 & 5.43 & & & \\
224297601 & 271.0980 & -24.4987 & y & n* & {0} & P & 20.08 & 17.86 & 16.42 & 15.73 & 16.06 & K7 & 0.47 & 0.91 & 0.75 & 1.86 & 6.17 & & & \\
224305947 & 271.1103 & -24.4834 & y & y & {0} & QPD & 19.44 & 17.06 & 15.49 & 14.77 & 15.16 & K7.3 & 1.15 & 3.93 & 0.61 & 3.93 & 5.15 & & & \\
224308921 & 271.0940 & -24.4781 & y & y* & {1} & QPS & 16.31 & 15.10 & 14.48 & 14.16 & 14.35 & F5.3 & 1.02 & 4.87 & 1.38 & 1.73 & 7.14 & -8.28 & 0.26 & \\
224311070 & 270.9652 & -24.4740 & y & y & {1} & APD & 16.88 & 15.40 & 14.49 & 14.07 & 14.17 & G6 & 1.11 & 5.30 & 1.93 & 2.51 & 6.57 & -8.08 & 0.22 & 0.72 \\
224311432 & 271.1643 & -24.4734 & ? & n & {1} & N & 15.81 & 14.59 & 13.93 & 13.64 & 13.67 & G4.5 & 0.34 & 4.72 & 1.80 & 2.30 & 6.69 & -8.27 & 0.46 & 0.28 \\
224311447 & 271.1992 & -24.4733 & y & n & {0} & MP & 11.56 & 11.27 & 11.26 & 11.26 & 11.21 & A1.5 & 0.37 & 79.5 & 2.76 & 3.51 & 6.42 &  &  &  \\
224312151 & 271.0848 & -24.4721 & y & n & {2} & QPS & 17.76 & 15.71 & 14.36 & 13.77 & 13.91 & K7 & 0.25 & 5.11 & 0.63 & 4.42 &  & -7.40 & 0.19 & 0.16 \\
224313065 & 271.1710 & -24.4704 & ? & n & {1} & N & 15.50 & 14.39 & 13.91 & 13.72 & 13.77 & G1.3 & 0.22 & 4.46 & 1.61 & 2.10 & 6.86 & -8.10 & 0.16 & 0.17 \\
224314197 & 270.8580 & -24.4683 & y & n & {0} & P & 18.33 & 16.43 & 15.40 & 14.99 & 15.18 & K3.5 & 0.31 & 1.55 & 1.23 & 1.94 & 6.37 &  &  & \\
\enddata
\tablecomments{A full version of this table is provided in electronic format. A portion is shown here for guidance regarding its form and content.}
\tablenotetext{a}{Cross-identification number from one of the following catalogs: Ecliptic Plane Input Catalog (EPIC; \citealp{huber2016}); \citet[][SCB]{Sung2000}; \citet[][WFI]{prisinzano2005}; 2MASS; VPHAS+; GES.}
\tablenotetext{b}{Disk classification as ``y'' (high-confidence disk-bearing), ``y*'' (disk-bearing with more uncertain classification), ``n'' (high-confidence disk-free), ``n*'' (disk-free with more uncertain classification), ``?'' (disk candidate).}
\tablenotetext{c}{Accreting status classification as ``1'' (accreting), ``2'' (potential accretor), ``0'' (non-accreting).}
\tablenotetext{d}{See \citet{Venuti2021}; possible values include ``P'' (periodic), ``QPS'' (quasi-periodic symmetric), ``S'' (stochastic), ``B'' (burster), ``QPD'' (quasi-periodic dipper), ``APD'' (aperiodic dipper), ``N'' (flat-line), and ``U'' (unclassifiable).}
\tablenotetext{e}{Logarithmic amplitude of $\dot{M}_{acc}$ variations above the typical (median) $\dot{M}_{acc}$ value measured over the monitored period.}
\tablenotetext{f}{Logarithmic amplitude of $\dot{M}_{acc}$ variations below the typical (median) $\dot{M}_{acc}$ value measured over the monitored period. This parameter is only provided for objects that showed detectable $\dot{M}_{acc}$ levels at all epochs during our monitoring campaign.}
\end{deluxetable*}

In order to assess the membership status of stars in our sample, we reviewed membership criteria available in the literature, which fall broadly into three categories:
\begin{itemize}
\item \textit{photometric}, which encompasses all emission properties indicative of a YSO nature, {including} X-ray detection, UV excess, IR excess, shape of the SED, and H$\alpha$ emission \citep[membership data from][]{Sung2000,Rauw2002,Damiani2004,Kumar2010,Povich2013,Broos2013,Castro2013,Kalari2015}; 
\item \textit{kinematic}, which encompasses kinematic or astrometric evidence of belonging to the cluster {from} radial velocity, proper motion, and/or parallax data \citep{vandenancker1997,Chen2007,Prisinzano2007,Damiani2019}; 
\item \textit{spectroscopic}, {i.e.,} Li absorption in the stellar spectrum \citep{Prisinzano2007,Arias2007}.
\end{itemize}

For each object in our list, we flagged membership according to individual criteria, and then assigned a final classification to our stars as either \textit{bona fide} members or candidate members. \textit{Bona fide} members were selected as objects displaying Li absorption indicative of youth, or satisfying at least one membership criterion from two different categories, or two distinct membership criteria from a single category. Candidate members were defined as objects with no recorded Li absorption measurement, and satisfying only one criterion for membership, or with member status originating from a single literature study. Results from this classification are reported in Table~\ref{tab:sample_properties}; 787/1012 stars were classified as \textit{bona fide} members (78\%), while the remaining 225 were classified as candidate members.

\subsection{Disk classification} \label{sec:disk_class}

As mentioned in Sect.~\ref{sec:auxiliary_datasets}, the available optical photometry was combined with literature IR photometric catalogs to classify the evolutionary status of individual YSOs with respect to their disk properties. The same multi-indicator classification procedure outlined in \citet{Venuti2021} was uniformly extended to the complete sample of 1012 stars considered here. In summary, the criteria considered for disk classification included: 
\begin{enumerate}
    \item[i.] the $\alpha_{IRAC}$ index defined at $[3.6]-[8.0]$ wavelengths \citep{Teixeira2012}; 
    \item[ii.] the location of individual stars on near-IR (2MASS $J,H,K$) and mid-IR (WISE $W1,W2,W3$) color-color diagrams with respect to the loci of disk-bearing stars \citep{Meyer1997,Koenig2014}; 
    \item[iii.] the reddening-free optical-infrared indices $Q_{JHHK}$ and $Q_{VIJK}$ \citep{Damiani2006}.
\end{enumerate}
Disk flags from the various disk indicators were then merged to assign a final disk class to each target. In case of consistent disk classification for a given source across all indicators, the corresponding class was assigned as final disk class. In case of discordant information from different indicators, we adopted the following approach:
\begin{itemize}
\item in the case of disk detection at mid-IR wavelengths but not at near-IR wavelengths, the target was classified as disk-bearing, as it may be hosting a more evolved disk, no longer traceable in the near-IR;\looseness=-1
\item in the case of disk detection at near-IR wavelengths but not at mid-IR wavelengths, the target was classified as a disk candidate object {if the mid-IR flux measurements were still consistent with the disk-bearing color loci within the photometric uncertainties, and as disk-free otherwise};
\item in the case of a disk candidate flag in the mid-IR and a disk non-detection in the near-IR, the object was retained as a disk candidate;
\item in the case of a disk non-detection in the mid-IR and a disk candidate flag in the near-IR, the object was classified as disk-free.
\end{itemize}
This procedure enabled assignment of a disk class to 844/1012 objects. Among these, {229 (27.1\%)} were classified as disk-bearing, {568 (67.3\%)} were classified as disk-free, and {47 (5.6\%)} were classified as disk candidate objects. For stars classified as either disk-bearing or disk-free, we also introduced a further flag aimed at qualifying the strength of the disk status classification. For disk-free stars, we attributed a strong classification (high-confidence) flag to sources for which at least two distinct disk indicators in separate IR band sets were available, with all providing a coherent no-disk classification. For disk-bearing stars, we attributed a strong classification flag to sources that satisfied either of the following criteria: i) $\alpha_{IRAC} > -1.8$, corresponding to a thick disk classification; ii) availability of disk indicators in all three band sets, and definite disk-bearing status according to at least two of them; iii) availability of disk indicators in two separate band sets, and classification incompatible with disk-free status at all available bandpasses; iv) availability of multiple disk indicators in a given band set (namely $J,H,K$), and definite disk-bearing classification from all indicators. This additional flagging system resulted in $\sim$73\% of disk-free stars and $\sim$64\% of disk-bearing stars being assigned a strong classification label, while sources that did not satisfy the listed criteria were considered more uncertain disk-bearing/disk-free classifications. Results from this classification are reported in Table~\ref{tab:sample_properties}.

To derive a statistical estimate of the true disk fraction in our sample, which accounts for objects with no disk status information, we {explored a grid of theoretical disk fractions extending over the entire range of possible values with a step of 0.01, and for each step of the grid we simulated a population of 1012 stars with an injected disk fraction corresponding to the grid value.} 
We then extracted 844 members among the simulated sample of 1012, and counted for how many of those a disk-bearing status had been assigned. For each injected disk fraction, {we repeated this procedure for 10\,000 times, and out of these 10\,000 iterations} we tallied the instances for which the number of ``observed'' stars with disk-bearing status, extracted from the total simulated sample, was consistent with the real observations (i.e., comprised between a minimum dictated by the number of stars classified as disk-bearing with high confidence in our sample, and a maximum dictated by the sum of disk-bearing and disk candidate objects, plus low-confidence disk-free objects). {We then built a distribution of the resulting tallies as a function of the assumed disk fraction, and} fitted {its profile} with a Gaussian distribution, whose mean and standard deviation provided us with a best statistical disk fraction $\mathit{f}_{disk}$, estimated to be ${0.34 \pm 0.14}$ across our entire sample. When restricting this statistical analysis to \textit{bona fide} members only, a similar $\mathit{f}_{disk} = {0.37 \pm 0.16}$ was derived.

\section{Stellar properties}\label{sec:stellar_properties}

In this section, we describe how stellar parameters were estimated from our collection of multi-band photometry. We start with individual A$_V$--SpT values (Sect.~\ref{sec:SpT_Av}), which are required, together with suitable effective temperature (\teff; Sect.~\ref{sec:Teff}) and bolometric correction (BC; Sect.~\ref{sec:BC}) scales, to derive bolometric luminosities (\lbol; Sect.~\ref{sec:Lbol}). We then build the Hertzsprung-Russell (HR) diagram of the cluster, and estimate stellar masses, ages, and radii (Sect.~\ref{sec:HR_diagram}). All parameters obtained for each star are reported in Table~\ref{tab:sample_properties}, while the associated uncertainties are listed as a function of spectral class in Table~\ref{tab:param_uncertainties}. In turn, the derived stellar parameters are key to calculating mass accretion rates, as presented later in Sect.~\ref{sec:Mdot}.

\begin{deluxetable}{c c c c c c c c}
\centerwidetable
\tablecaption{Typical uncertainties on the stellar and accretion parameters derived here as a function of spectral class. \label{tab:param_uncertainties}}
\tablehead{
 \colhead{SpT} & \colhead{$\sigma_{T_\mathrm{eff}}$} & \colhead{$\sigma_{BC_V}$} & \colhead{$\sigma_{M_\mathrm{bol}}$} & \colhead{$\sigma_{L_\mathrm{bol}}$} & \colhead{$\sigma_{M_\star}$} & \colhead{$\sigma_{R_\star}$} & \colhead{$\sigma_{\log{\dot{M}}}$}\\
 & [$T_\mathrm{eff}$] & [mag] & [mag] & [$L_\mathrm{bol}$] & [$M_\star$] & [$R_\star$] & [dex]} 
\startdata
O & 0.020 & 0.25 & 0.40 & 0.37 & -- & 0.19 & --\\
B & 0.085 & 0.15 & 0.34 & 0.32 & 0.11--0.14 & 0.23 & 0.34\\
A & 0.015 & 0.14 & 0.34 & 0.31 & 0.10--0.12 & 0.16 & 0.34\\
F & 0.020 & 0.08 & 0.32 & 0.29 & 0.11--0.13 & 0.15 & 0.34\\
G & 0.035 & 0.09 & 0.32 & 0.30 & 0.16--0.18 & 0.17 & 0.36\\
K & 0.050 & 0.08 & 0.32 & 0.29 & 0.25--0.32 & 0.18 & 0.38\\
M & 0.030 & 0.26 & 0.40 & 0.37 & 0.18--0.23 & 0.19 & 0.38\\
\enddata
\tablecomments{The typical estimates provided for $\sigma_{T_\mathrm{eff}}$, $\sigma_{L_\mathrm{bol}}$, $\sigma_{M_\star}$, and $\sigma_{R_\star}$ are fractional uncertainties.
No typical $\sigma_{M_\star}$ range and $\sigma_{\log{\dot{M}}}$ are reported for O-type stars because no $M_\star$ estimates could be derived for them following the method outlined in the text.}
\end{deluxetable}

\subsection{Individual extinctions and spectral types} \label{sec:SpT_Av}

Estimates of SpT were derived for our targets by following the procedure described in \citeauthor{Venuti2021} (\citeyear{Venuti2021}; {see their Fig.~3 for an illustration of the method}). In a nutshell, the observed colors of each star on the ($r-i$, $g-r$) diagram were compared to the unreddened reference SpT--color sequence tabulated in VST/OmegaCAM filters by \citet{drew2014}. For each spectral subclass, we calculated the required color corrections (i.e., A($g-r$) and A($r-i$)) to be applied to our observed colors in order to match the expected colors for that SpT, assuming the anomalous reddening law reported by \citet{prisinzano2019}. We then selected the SpT estimate that corresponded to the best agreement between the calculated A($g-r$) and A($r-i$), and defined our best A$_V$ as the average between the derived, color-dependent extinction coefficients.\looseness=-1

In order to strengthen the SpT determination, we also applied the same procedure to the ($V-I$, $B-V$) diagram built from literature photometry, using the reference sequence of unreddened colors tabulated by \citet{pecaut2013}. Although potentially affected by larger uncertainties due to data non-homogeneity and non-simultaneity, the latter diagram benefited our A$_V$--SpT derivation procedure by mitigating the impact of degenerate trends between the ($r-i$, $g-r$) color sequence and the adopted reddening law that prevented accurate SpT determinations for some objects, as reported in \citet{Venuti2021}. The individual solutions from each diagram were then merged as follows:
\begin{enumerate}
    \item if the best solutions from both diagrams were consistent within half a spectral class, an averaged solution was adopted as best A$_V$--SpT parameters;
    \item if multiple valid solutions emerged from one or both diagrams, we selected the one that provided the best agreement between the two sets of results;
    \item if no converging solution could be found, and in the absence of literature information suggesting otherwise, we favored the result extracted from $g,r,i$ colors, {owing to their simultaneity and widespread availability across} our sample of Lagoon Nebula members (while $B,V,I$ colors are only available for 63\% of our targets).
\end{enumerate}

At the end of this procedure, we were able to assign an SpT estimate to 966/1012 (95\%) stars in our sample, around 51\% of which had individual SpT estimates from both sets of $g,r,i$ and $B,V,I$ colors. The resulting spectral class distribution across our sample is as follows:
\begin{itemize}
    \itemsep0em
    \item[--] O-type stars: 0.7\%
    \item[--] B-type stars: 4.6\%
    \item[--] A-type stars: 8.0\%
    \item[--] F-type stars: 11.6\%
    \item[--] G-type stars: 14.2\%
    \item[--] K-type stars: 50.4\%
    \item[--] M-type stars: 10.5\%
\end{itemize}

\subsection{Effective temperatures (\teff)} \label{sec:Teff}

The derived SpT were converted to \teff{} following a literature reference scale. Multiple published SpT--$T_\mathrm{eff}$ scales were compared, and their class-by-class difference evaluated, to estimate the uncertainty resulting from this conversion on the SpT-dependent $T_\mathrm{eff}$ values. More specifically, we considered the following scales:
 
\begin{itemize}
 \item \citet{bohm-vitense1981}, \citet{pecaut2013}, and \citet{gray2009} for O-type stars;
 \item \citet{bohm-vitense1981}, \citet{pecaut2013}, \citet{gray2009}, and \citet{kenyon1995} for B-type stars;
 \item \citet{gray2009}, \citet{pecaut2013}, and \citet{kenyon1995} for A-type stars;
 \item \citet{gray2009}, \citet{pecaut2013}, \citet{herczeg2014}, and \citet{kenyon1995} for F/G/K stars;
 \item \citet{gray2009}, \citet{pecaut2013}, \citet{herczeg2014}, and \citet{Luhman2003} for M-type stars.
\end{itemize}
 
For each compilation and spectral subclass, we calculated the absolute difference in tabulated $T_\mathrm{eff}$ with respect to \citeauthor{gray2009}'s (\citeyear{gray2009}) scale (which provides the most extensive SpT coverage). The largest percentage discrepancy measured across all scales for each subclass was then selected, and the final fractional uncertainty on $T_\mathrm{eff}$ was defined class by class from the median percentage error across all relevant subclasses, rounded to the closest multiple of 0.5\%. These uncertainties are reported in Table~\ref{tab:param_uncertainties}, while individual $T_\mathrm{eff}$ values were assigned using \citeauthor{gray2009}'s (\citeyear{gray2009}) SpT--$T_\mathrm{eff}$ scale.

\begin{figure}
\centering
\includegraphics[width=0.47\textwidth]{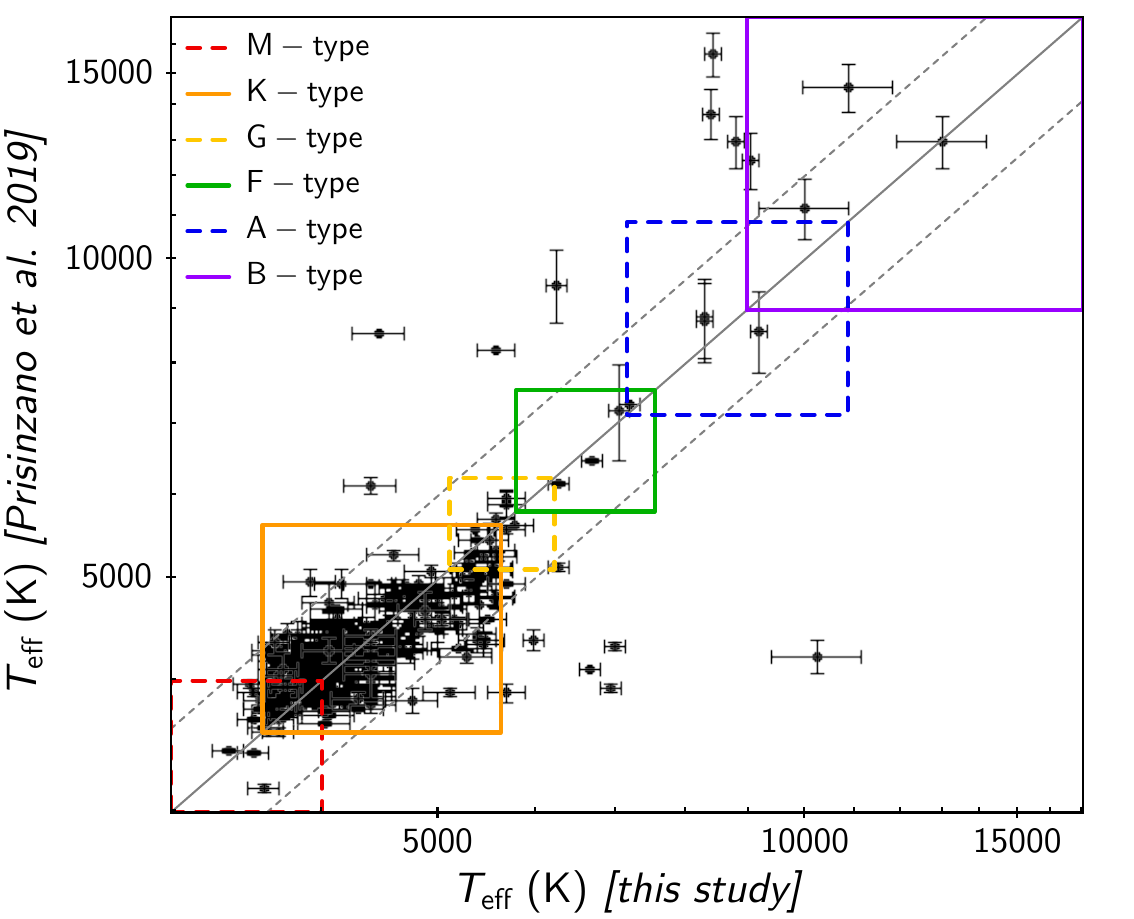}
\caption{{Comparison between \teff{} estimates derived in this study as presented in Sect.~\ref{sec:Teff} and the values obtained in \citet{prisinzano2019} for objects in common between the two samples. The solid gray line traces the equality line on the diagram, while the dotted gray lines trace a difference between the two \teff{} sets that amounts to a factor of 1.2. The colored boxes mark the ranges in $T_\mathrm{eff}$ associated with different spectral classes, as labeled in the legend. The overlapping corners between adjacent boxes correspond to the transition between ninth subclass of the previous spectral group and zeroth subclass of the following spectral group, while the delimiting $T_\mathrm{eff}$ values take the uncertainties listed in Table~\ref{tab:param_uncertainties} into account.}}
\label{fig:Teff_comp}
\end{figure}

Independent, spectroscopically--determined \teff{} values were obtained for $\sim$34\%  of our sample within the Gaia-ESO Survey \citep[GES;][]{gilmore2012,randich2013}, as published in \citet{prisinzano2019}. A comparison between our adopted \teff{} values and the GES parameters revealed that the vast majority (90\%) of objects in common have consistent \teff{} estimates within a factor $\sim$1.2, {corresponding to a range of 1--3 spectral subclasses, both at the lower mass end (G/K stars) and at the higher mass end (A stars)}, {as illustrated in Fig.~\ref{fig:Teff_comp}, and only $\sim$4\% of objects appear as significant outliers with inconsistent spectral classes assigned from the two analyses.} 
{Disk--bearing and disk--free stars exhibit a similar distribution across the diagram, with median measured ratios between our $T_\mathrm{eff}$ values and \citeauthor{prisinzano2019}'s (\citeyear{prisinzano2019}) estimates that amount to 1.00 and 0.98, respectively. The interquartile range measured around the central value for disk--bearing stars (0.13) is slightly larger than that measured for disk-free stars (0.10), possibly reflecting an additional source of scatter associated with some residual contribution of disk--related phenomena to the optical stellar colors. Nevertheless, the overall consistency in stellar properties between disk--bearing and disk--free stars that we observe at $g,r,i$ wavelengths indicates that this effect, while it may reduce the accuracy of individual parameters derived for some stars with enhanced disk activity, would not impact the statistical inferences that we draw from our analysis.}

\subsection{Bolometric corrections} \label{sec:BC}

To convert our apparent magnitudes to bolometric magnitudes, we compiled a \teff{}--dependent $V$-band BC (BC$_V$) scale, with uncertainties, by comparing the individuals scales published by \citet{pecaut2013}, \citet{Bessell1998}, \citet{schmidt-kaler1982}, and \citet{kenyon1995}.
To derive a functional form that would link the values of $T_\mathrm{eff}$ and BC$_V$, we conducted a polynomial fit to each scale. 
We then compared the predictions between each pair of scales over the most restrictive $T_\mathrm{eff}$ range covered by both, and measured the absolute difference in predicted BC$_V$ values as a function of $T_\mathrm{eff}$. We used the $T_\mathrm{eff}$--SpT scale from \citet{gray2009} to separate the $T_\mathrm{eff}$ grid into spectral classes, and for each spectral class and pair of conversion scales we derived the typical uncertainty on BC$_V$ as the median discrepancy in predicted values at a given temperature, across the entire spectral class. We then defined the class-by-class uncertainty on BC$_V$ as the median difference calculated for each spectral class across all pairs of conversion scales. The resulting uncertainties on BC$_V$ are listed in Table~\ref{tab:param_uncertainties}. 

Our adopted BC$_V$ values were extracted from \citet{schmidt-kaler1982}, because it provides the most comprehensive coverage in spectral types among the references considered here. 

\subsection{Bolometric magnitudes and luminosities} \label{sec:Lbol}

In order to apply the BC$_V$ scale to optical magnitudes in our sample, we used the transformation relationships by \citet{jester2005} and \citet{jordi2006} to convert the VST/OmegaCAM measurements to Johnson-Cousins $V$-band magnitudes. Both transformations use the observed $g$-band magnitudes and $g-r$ colors, and yield consistent results within $\sim$0.009~mag. On the other hand, {the typical absolute difference between converted and archival $V$-band magnitudes (when available) across our sample ranges from $\sim$0.03~mag at $V\sim11.2$ to $\sim$0.09~mag at $V\sim18.8$, which encompass 90\% of the population over which this comparison could be performed. We then assumed the median value} $\sim$0.06~mag as uncertainty on the derived $V$-band magnitudes, averaged from the two sets of transformations. Absolute magnitudes $M_V$ were then derived as $V-A_V-5\cdot(\log{d}-1)$, with an adopted distance $d$ of $1325 \pm 113$~pc \citep{Damiani2019}. \teff{}--dependent $BC_V$ values were added to the calculated $M_V$ to derive the bolometric magnitudes $M_\mathrm{bol}$, and bolometric luminosities $L_\mathrm{bol}/L_\odot$ were estimated as $10^{0.4\cdot(M_{\mathrm{bol},\odot}-M_\mathrm{bol})}$.

To account for the uncertainties discussed earlier, we estimated the following typical parameter errors:
\begin{itemize}
\item $\sigma_{M_V} = \sqrt{\sigma^2_V + \sigma^2_{A_V} + \left[\left(\frac{5}{d \cdot \ln{10}}\right)\sigma_{d}\right]^2} \simeq 0.31$, following from the aforementioned uncertainties on $V$\footnote{{We note that assuming a magnitude--dependent $\sigma_V$, as opposed to the uniform value discussed earlier in the text, would lead to only a modest ($\sim$2\%) change in the derived $\sigma_{M_V}$ with respect to the typical estimate listed here.}} and $d$, and from the typical total uncertainty on the adopted value of $A_V$;
\item $\sigma_{M_\mathrm{bol}} = \sqrt{\sigma^2_{M_V} + \sigma^2_{BC_V}}$, dependent on spectral class as listed in Table~\ref{tab:param_uncertainties};
\item $\frac{\sigma_{L_\mathrm{bol}}}{{L_\mathrm{bol}}} = 0.4\cdot\ln{10}\cdot\sigma_{M_\mathrm{bol}}$, also dependent on spectral class, as reported in Table~\ref{tab:param_uncertainties}.
\end{itemize}

\subsection{Stellar masses ($M_\star$), ages ($t$), and radii ($R_\star$)} \label{sec:HR_diagram}

\begin{figure}
\centering
\includegraphics[width=0.47\textwidth]{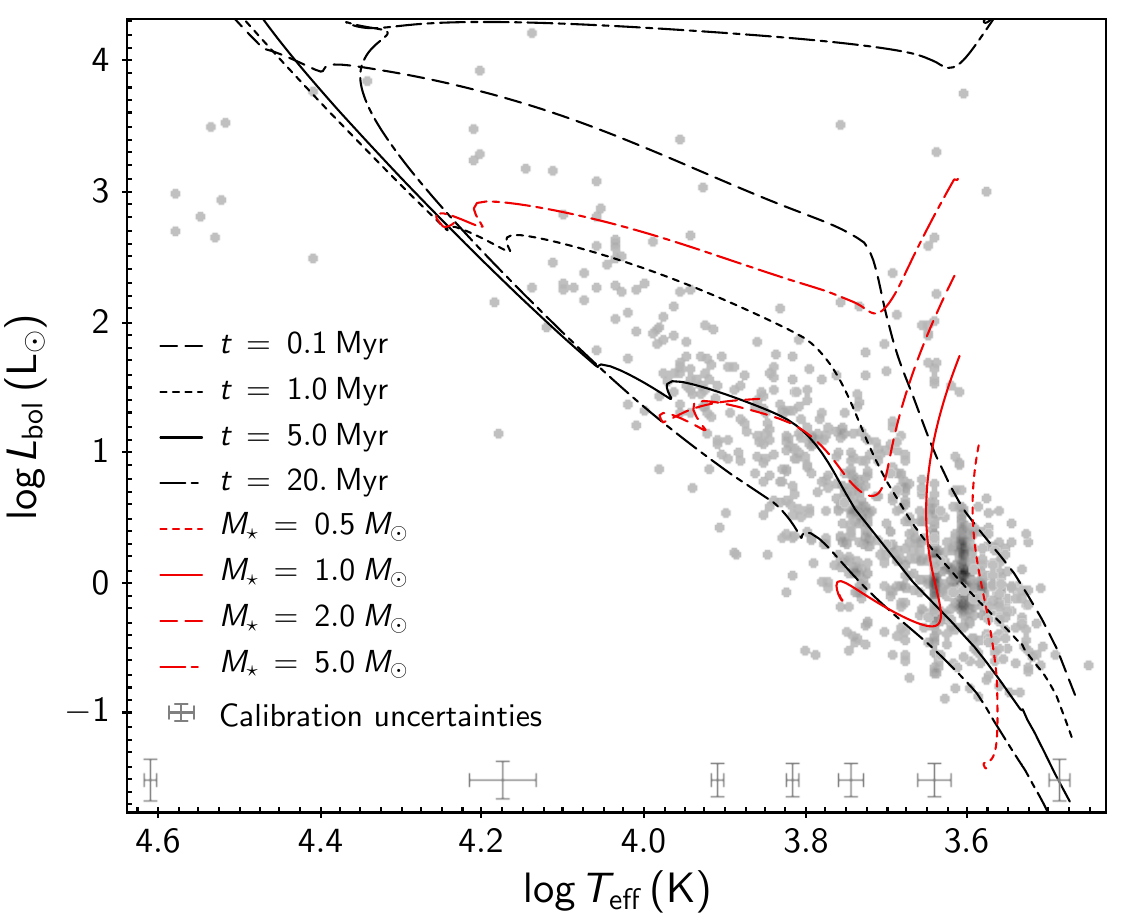}
\caption{Distribution of Lagoon Nebula members on the HR diagram. Model mass tracks (red) and isochrones (black) shown on the diagram were extracted from the MIST compilation \citep{dotter2016, Choi2016}, assuming solar metallicity and with no rotation effects. The error bars at the bottom illustrate the SpT-dependent typical uncertainties on $T_\mathrm{eff}$ and $L_\mathrm{bol}$ values, as discussed in Sects.~\ref{sec:Teff} and \ref{sec:Lbol}.}
\label{fig:Lagoon_HR_diagram_MIST_nonrot_solar}
\end{figure}

The derived values of $T_\mathrm{eff}$ and $L_\mathrm{bol}$ were used to build the HR diagram of the region, illustrated in Fig.~\ref{fig:Lagoon_HR_diagram_MIST_nonrot_solar}.
\citet{david2019} presented a thorough comparative assessment of different pre-main sequence evolutionary model grids with regards to the determination of intrinsic stellar parameters. Based on their analysis, here we adopt the MESA (Modules for Experiments in Stellar Astrophysics) Isochrones \& Stellar Tracks (MIST) models \citep{dotter2016, Choi2016}, without rotation and with solar metallicity, to interpolate the position of our targets on the HR diagram and estimate individual masses ($M_\star$) and ages ($t$). For each object, we extracted the ($\log{T_\mathrm{eff}}$, $\log{L_\mathrm{bol}}$) coordinates of the delimiting points from the two closest isochrones and mass tracks around its position, and conducted a linear interpolation over $\log{L_\mathrm{bol}}$ and $\log{T_\mathrm{eff}}$ to extract best-fit $t$ and $M_\star$ values. The corresponding $M_\star$ uncertainty ($\sigma_{M_\star}$) was then estimated by drawing a $[\pm\sigma_{T_\mathrm{eff}},\pm\sigma_{L_\mathrm{bol}}]$ box around the datapoint on the diagram, and determining the limiting $M_\star$ values from the range of isochrones that intersected the boxed area. The resulting typical $\sigma_{M_\star}$ ranges are listed as a function of SpT in Table~\ref{tab:param_uncertainties}.

Overall, an estimate of $M_\star$ was obtained for 924/1012 targets in our sample (91\%); {among the remaining 88 objects, 40 have no (\lbol{}, \teff{}) derivations for placement onto the HR diagram, 43 appear located below the main sequence turnoff on Fig.~\ref{fig:Lagoon_HR_diagram_MIST_nonrot_solar}, and 5 appear located above the pre-main sequence onset (i.e., brighter than predicted for the earliest ages at the corresponding \teff{}) on the diagram. Disk--bearing stars appear to be overrepresented among the sources with no $M_\star$ estimate (29 vs. 30 disk--free stars, in spite of the fact that disk--free stars are 2.5 times as frequent as disk--bearing stars in our sample, as discussed in Sect.~\ref{sec:disk_class}). In both groups, the subsets of objects with no $M_\star$ are about evenly split between cases lacking (\lbol{}, \teff{}) estimates and cases for which no isochrone--fitting solution could be extracted due to their position on the HR diagram. However, among the disk--free cases, only about one third of objects with no isochrone--fitting solution have a strong no-disk classification, which suggests the apparent properties of HR diagram outliers with respect to the pre-main sequence model grid may be linked to some contamination by disk--related phenomena.} 

\begin{figure}
\centering
\includegraphics[width=0.47\textwidth]{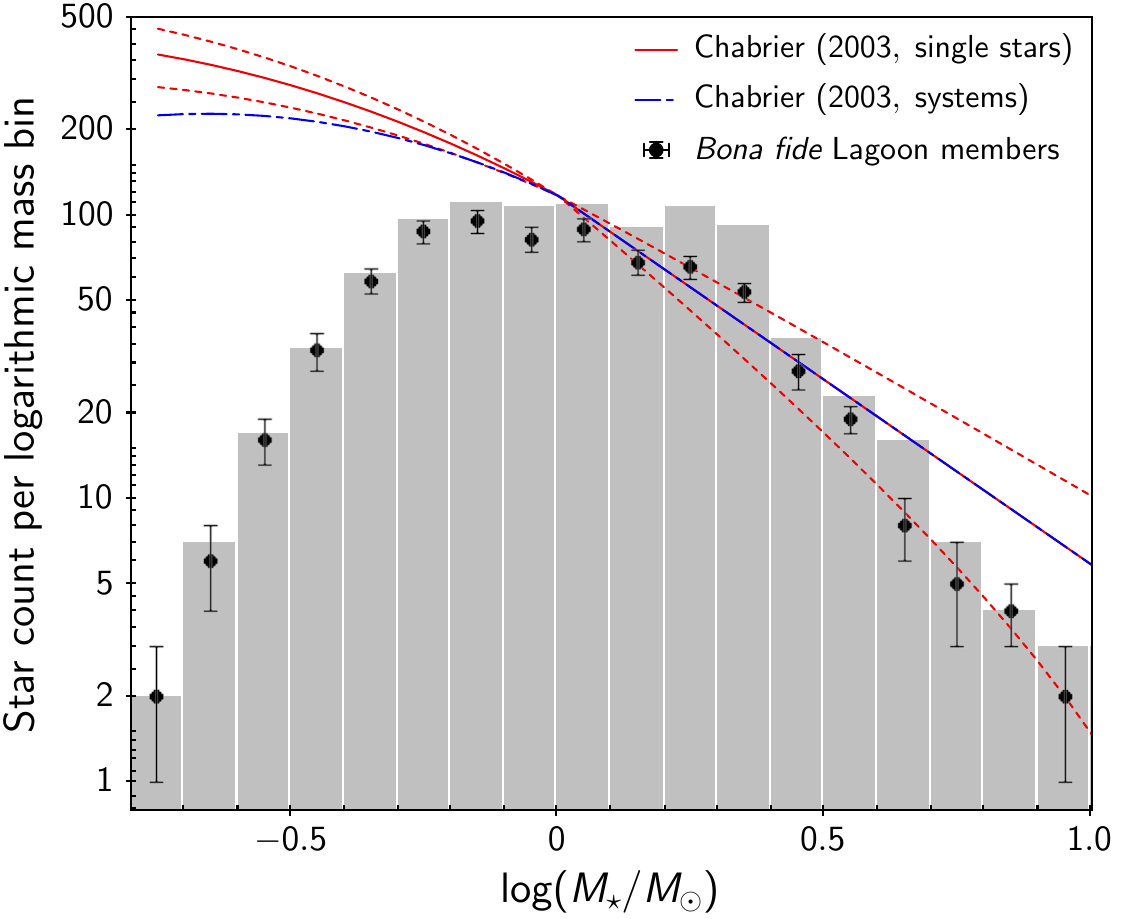}
\caption{Mass distribution inferred for young stars in our Lagoon Nebula sample. The gray histogram refers to the entire population considered in this study, while the black dots and corresponding uncertainties trace the distribution of \textit{bona fide} members only. The IMF expressions presented in \citet{chabrier2003} for single stars (red solid line) and stellar systems (blue dash-dotted line) in the Galactic field are overlaid for comparison purposes, renormalized to the peak of the observed distribution. The red dotted lines trace the range of possible values around the single--star IMF ensuing from the uncertainties on the function parameters tabulated by \citet{chabrier2003}.}
\label{fig:mass_distribution}
\end{figure}

The mass distribution derived across our sample is illustrated in Fig.~\ref{fig:mass_distribution}.
The 1\textsuperscript{st}--99\textsuperscript{th} percentiles range in $M_\star$ extends from $\sim0.25-6.45$\,$M_\odot$, encompassing both the T~Tauri mass regime ($M_\star \lesssim 2$\,$M_\odot$) and the Herbig Ae/Be mass regime ($M_\star \sim 2-10$\, $M_\odot$; e.g., \citealp{bastien2015}).
To evaluate our completeness levels across the stellar mass spectrum, we compared our data to the initial mass function (IMF) expressions presented by \citet{chabrier2003} for single stars and multiple system populations in the Galactic disk. Earlier studies, focused on different star forming regions \citep[e.g.,][]{moraux2005} and on the Lagoon Nebula itself \citep{prisinzano2005}, have shown an overall agreement between the shape of the observed mass distributions in young clusters and the predictions based on the Galactic field IMF. Such an agreement can be seen in Fig.~\ref{fig:mass_distribution} for masses $M_\star > 1 M_\odot$. While the match in absolute value derives from the arbitrary normalization of the IMF to the observed distribution peak, the correspondence between the IMF curve and the declining trend of star count with increasing mass in this range indicates that, at least on a relative scale, the mass spectrum of our sample adequately represents the expected composition of the Lagoon Nebula population. At lower masses, however, our sample appears significantly incomplete, with an estimated fraction of missing objects $\sim15
\%$ at $M_\star \sim 0.85\, M_\odot$, $\sim50
\%$ at $\sim 0.55\, M_\odot$, and $\sim90
\%$ at $\sim 0.3\, M_\odot$. These biases have been taken into account when evaluating mass--dependent trends in disk activity across the cluster, as discussed in the following sections. {The mass distribution of disk--bearing members peaks at smaller values than the center of the distribution measured for disk--free members: the median $\log{M_\star}$ and the associated interquartile range amount respectively to $-0.04$ and 0.41 in the first case, and to $+0.10$ and 0.50 in the second case. This difference between the two mass distributions may reflect distinct disk evolutionary stages for stars of different mass, as examined in Sect.~\ref{sec:facc_fdisk_mass}.}

The median age extracted from the MIST models across the population is $\log{t} = 6.3_{-0.5}^{+0.4}$ (or $\log{t} = 6.2 \pm 0.4$ for \textit{bona fide} members only), slightly older but consistent with the typical age of $\lesssim 1$~Myr associated with the NGC~6530 cluster \citep{prisinzano2019}. {We note, however, that the average age $\log{t} = 5.84 \pm 0.36$ reported by \citet{prisinzano2019} for NGC~6530 is limited to solar-mass and low-mass cluster members ($T_\mathrm{eff}<5500\,K$); if we applied the same cuts to our member sample, we would obtain a more similar $\log{t} = 6.0_{-0.3}^{+0.4}$. This discrepancy may be due to systematic trends in model--based age predictions for higher--mass (hotter) stars with respect to lower--mass (cooler) stars on the pre-main sequence \citep[e.g.,][]{hosokawa2011,herczeg2015}. Disk--bearing stars appear slightly younger than the cluster average, with a median $\log{t} = 6.1_{-0.3}^{+0.5}$ (or $\log{t} = 6.0_{-0.3}^{+0.5}$ among \textit{bona fide} disk--bearing members only). This feature, coupled with the larger proportion of disk--free stars in our sample, could also contribute to explaining the slightly older age we measure here for the Lagoon Nebula with respect to \citeauthor{prisinzano2019}'s (\citeyear{prisinzano2019}) estimate, which was derived over a population where disk--bearing/accreting stars were predominant. The datapoint distribution in Fig.~\ref{fig:Lagoon_HR_diagram_MIST_nonrot_solar} suggests a considerable age spread among cluster members, the extent of which (spanning over 2~dex) overlaps with the age ranges previously reported in the literature \citep{tothill2008,prisinzano2019}. Compared to \citeauthor{prisinzano2019}'s (\citeyear{prisinzano2019}) results, we find here a larger proportion of older ($>$5~Myr) sources, albeit with a strong mass--dependent distribution: among $M_\star \leq 1\,M_\odot$ sources, 87\% appear younger than 5~Myr, while among sources more massive than this threshold the apparent percentage of $t<5$~Myr stars is only 61\%. Disregarding the mass dependence (which may be impacted by systematic effects in the model predictions), the subset of $M_\star \leq 1\,M_\odot$ that appear older than 5~Myr is characterized by a slightly larger proportion of candidate members (38\%) and a smaller proportion of high-confidence disk--free stars (19\%) compared to the the overall population. This would suggest that a true tail of old (up to $t\sim 15$~Myr; \citealp{prisinzano2019}) cluster members may be mixed with sources affected by photometric uncertainties and/or residual disk effects in this region of the HR diagram.} 

To complete our set of stellar parameters, stellar radii $R_\star$ were also estimated for our targets from the derived $T_\mathrm{eff}$ and $L_\mathrm{bol}$ using the Stefan-Boltzmann law.

\section{Tracers of accretion activity} \label{sec:accretion_tracers}

To trace accretion activity uniformly across our sample, we relied on the VST/OmegaCAM photometry and measured the color excess linked to accretion that can be detected above the photospheric emission level at UV wavelengths ($u$-band) and in the narrowband $H\alpha$ filter.

\subsection{$u-r$ color excess} \label{sec:E_ur}

\begin{figure}
\centering
\includegraphics[width=0.47\textwidth]{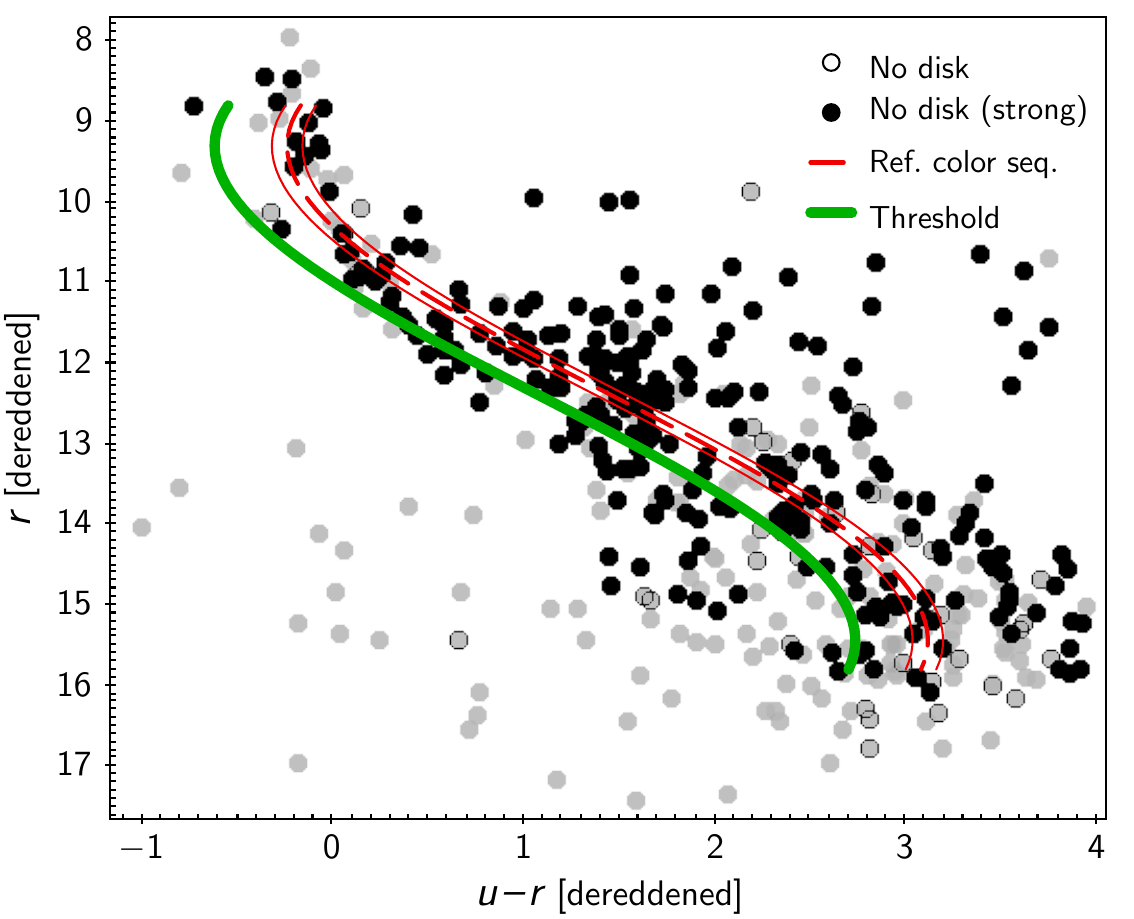}
\caption{Dereddened ($u-r$, $r$) CMD for Lagoon Nebula members in our sample with available $u$-band photometry. Objects classified as disk-free are highlighted as black dots (strong classification) and black circles (more uncertain classification). Gray dots with no contour correspond to disk-bearing sources. The red dashed and solid curves trace respectively the median reference color sequence extracted from our high-confidence disk-free population, and a photometric range around the sequence ($\pm 0.08$~mag) that takes into account the uncertainty on the reference fit. The threshold adopted to select accreting stars is traced in green.
}
\label{fig:ur_r_diagram_ref_colors}
\end{figure}

As a first indicator of accretion, following \citet{venuti2014}, we measured the $u-r$ color excess \Eur{} on the ($u-r$, $r$) color-magnitude diagram (CMD). As mentioned in Sect.~\ref{sec:sample_selection}, this indicator is available for 486/1012 sources in our sample. We first used the individual $A_V$ estimates to deredden our photometric data, using filter-dependent extinction coefficients calculated from the prescription of \citet{cardelli1989} at the filters' effective wavelengths and assuming an anomalous reddening law with R$_V$~=~5 \citep{prisinzano2019}. The resulting dereddened CMD is illustrated in Fig.~\ref{fig:ur_r_diagram_ref_colors}.

In order to measure \Eur, we used the subsample of stars, \textit{bona fide} members classified as high-confidence disk-free sources, to extract a magnitude-dependent reference color sequence $(u-r)_\mathrm{ref}$ that would trace the photospheric and chromospheric emission level above which accretion activity can be detected. To minimize the impact of color outliers, prior to extracting the reference sequence we removed the few dozen overluminous sources with photometric coordinates that would correspond to presumed ages more than $2 \sigma$ distant from the best-fitting isochrone to the cluster locus. 
For the retained subsample, we extracted a reference, $r$-dependent sequence of $u-r$ colors by implementing a moving median routine in 1~mag-wide bins in $r$ with steps of 0.5~mag. Within each bin, we computed the median $r$ magnitude ($\widetilde{r}$) and associated $u-r$ color, and then performed a third-degree polynomial fit to the ($\widetilde{r}$, $\widetilde{u-r}$) sequence to extract a functional form for the reference $u-r$ value as a function of $r$.

Two different sources of uncertainty are associated with the derived reference color sequence: one component that reflects the discrepancy between the actual median photometric sequence and the fitting curve, which amounts to about $\pm 0.08$~mag; another component that reflects the intrinsic scatter of observed photometric properties for disk-free stars around the reference sequence, estimated to be about $\pm 0.37$~mag from the typical extent of the interquartile range in $u-r$ values within each magnitude bin. The quadratic sum of these two sources of uncertainty amounts to $\sim 0.38$~mag; therefore, we considered as high-probability accretors all targets with $u-r$ colors that stand at least 0.38~mag below the reference color sequence for disk-free stars.

We defined the excess $E(u-r) = (u-r)_\mathrm{obs} - (u-r)_\mathrm{ref}$, computed across the entire sample of stars irrespective of their disk classification. We then used a tiered system to classify our targets according to their accretion status, based on the measured $E(u-r)$:
\begin{itemize}
\item {\it non-accreting} when $E(u-r) > -0.08$ (sources to the right of the reference color strip on Fig.~\ref{fig:ur_r_diagram_ref_colors});
\item {\it potential accretors} if $-0.38 < E(u-r) < -0.08$ (sources located between the reference color strip and the lower end of the interquartile range that traces the color dispersion of disk-free sources);
\item {\it accreting} when $E(u-r) < -0.38$ (i.e., located below the first quartile associated with the disk-free color distribution, traced in green on Fig.~\ref{fig:ur_r_diagram_ref_colors}).
\end{itemize}

The above procedure resulted in the following statistical classifications:
\begin{itemize}
    \item 120 accreting sources (25\%), 96 potential accretors (20\%), and 270 non-accreting sources (55\%) across the entire sample;
    \item {41}\% accreting sources and {45}\% non-accreting sources {among all} disk-bearing stars ({48}\% accreting sources and {38}\% non-accreting sources when limited to stars that were classified as disk-bearing with high confidence);
    \item 16\% {(15\%)} accreting sources and 63\% {(64\%)} non-accreting sources among {all (high-confidence)} disk-free stars.
\end{itemize}

\subsection{$r-H\alpha$ color excess} \label{sec:E_rHa}

As a second indicator of accretion activity, we measured the excess \ErHa{} on the observed $r-H\alpha$ colors. This approach could be applied uniformly across our sample, as 97\% of our targets have available $r,i,H\alpha$ photometry, as reported in Sect.~\ref{sec:sample_selection}. As a first diagnostic tool, we used the ($r-i$, $r-H\alpha$) color-color diagram (CCD), matching earlier accretion studies conducted on young star clusters (e.g., \citealp{Kalari2015}).

\begin{figure*}
\centering
\includegraphics[width=\textwidth]{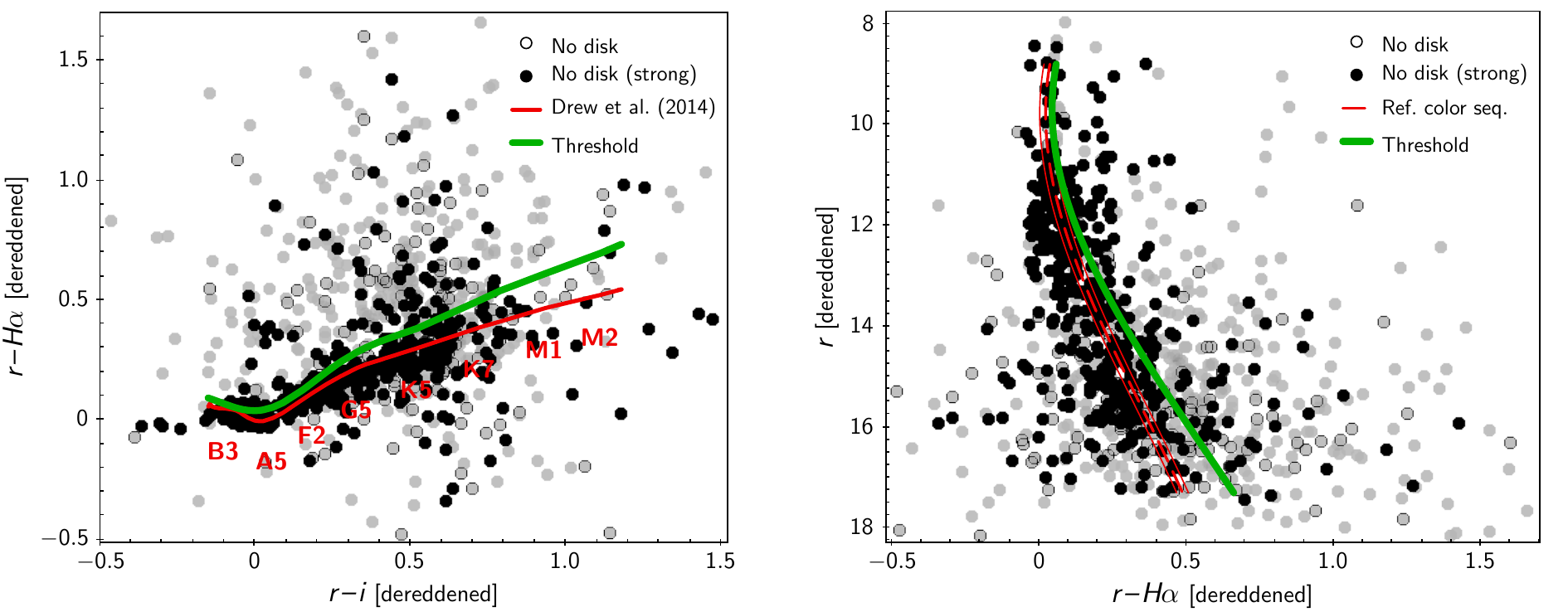}
\caption{Diagnostics of accretion extracted for our target stars from the $r-H\alpha$ color index on the dereddened ($r-i$, $r-H\alpha$) CCD (left) and ($r-H\alpha$, $r$) CMD (right). Colors and symbols are as in Fig.~\ref{fig:ur_r_diagram_ref_colors}. On the left panel, the red curve traces the SpT--dependent photospheric color sequence tabulated for main sequence dwarfs in VST/OmegaCAM filters by \citet{drew2014}. Spectral types, shifted downwards for clarity, are labeled along the reference sequence to guide the eye. On the right panel, the red dashed and solid curves trace respectively the median reference color sequence extracted from our high-confidence disk-free population, and a photometric range around the sequence ($\pm 0.02$~mag) that takes into account the uncertainty on the reference fit. On both panels, the threshold curve adopted to select accreting stars is traced with a thicker green line. {Stars selected as accretors are those located above the green line on the CCD, and to the right of the green line on the CMD.}}
\label{fig:ri_rHa_accretion_threshold}
\end{figure*}

The procedure we adopted to classify our targets as accreting or non-accreting based on $E(r-H\alpha)^{CCD}$ is illustrated in Fig.~\ref{fig:ri_rHa_accretion_threshold} (left).
Namely, we took as reference the color sequence in VST/OmegaCAM filters tabulated by \citet{drew2014}, which provides a very good description to the locus traced by the bulk of high-confidence disk-free stars on the diagram. We then defined a color-dependent accretion threshold as the sum of the adopted reference color sequence plus half the typical interquartile range in $r-H\alpha$ values among well-characterized disk-free stars as a function of $r-i$, which varies between $\sim 0.02$~mag at $r-i \simeq -0.15$ and $\sim 0.18$~mag at $r-i \simeq 1.17$. We further defined as potential accretors all stars that have measured $E(r-H\alpha)^{CCD}$ between 0.003~mag and the threshold curve, the lower limit being the typical uncertainty on the polynomial fit used to describe the reference color sequence with respect to \citeauthor{drew2014}'s (\citeyear{drew2014}) table. All remaining sources were considered non-accreting based on $E(r-H\alpha)^{CCD}$.

This procedure resulted in the following statistical classifications:
\begin{itemize}
    \item 385 accreting sources (39\%), 166 potential accretors (17\%), and 428 non-accreting sources (44\%) across the entire sample;
    \item 59\% (65\%) accreting sources and {26}\% (24\%) non-accreting sources {among all} (high-confidence) disk-bearing stars;
    \item 26\% ({21}\%) accreting sources and 54\% ({59}\%) non-accreting sources {among all} (high-confidence) disk-free stars.
\end{itemize}

We also conducted a classification of accreting vs. non-accreting sources based on their color distribution on the ($r-H\alpha$, $r$) CMD. The implementation of this procedure is illustrated in Fig.~\ref{fig:ri_rHa_accretion_threshold} (right).
Namely, as in Sect.~\ref{sec:E_ur}, we selected well-characterized disk-free stars and used them to build a reference, $r$--dependent color sequence above which to calculate the $r-H\alpha$ color excess $E(r-H\alpha)^{CMD}$. The reference sequence was derived by implementing a moving median technique on $r$ with the same binning choice used in Sect.~\ref{sec:E_ur}, and a third-order polynomial fit was extracted from the derived ($\widetilde{r}$, $\widetilde{r-H\alpha}$) sequence. The overall agreement between the fitting polynomial and the actual moving average sequence, as well as the typical interquartile range calculated for the spread in $r-H\alpha$ of disk-free stars as a function of $r$, were taken into account as described in Sect.~\ref{sec:E_ur} to derive the classification thresholds above the reference color sequence. A tiered classification scheme mirroring the one implemented in Sect.~\ref{sec:E_ur} was then applied to the entire population. A given object was classified as non-accreting if $E(r-H\alpha)^{CMD} < 0.02$, as a potential accretor if $E(r-H\alpha)^{CMD} = [0.02, 0.06]$, and as accreting if $E(r-H\alpha)^{CMD} > 0.06$.

Results from this classification are as follows:

\begin{itemize}
    \item 389 accreting sources (40\%), 123 potential accretors (12\%), and 467 non-accreting sources (48\%) across the entire sample;
    \item {58\% (62\%)} accreting sources and 30\% (28\%) non-accreting sources {among all} (high-confidence) disk-bearing stars;
    \item {31\% (28\%)} accreting sources and {56}\% (59\%) non-accreting sources {among all} (high-confidence) disk-free stars.
\end{itemize}

\subsection{Comparison between distinct accretion diagnostics} \label{sec:acc_class_comp}

The two ($r-H\alpha$)--based classifications schemes yield very similar proportions of accretors, $\sim40\%$ across the entire population, and $60\%-65\%$ among disk-bearing sources alone. The somewhat lower percentage of accreting stars selected with the \Eur{} indicator (25\% across the entire sample and up to $\sim$50\% among disk-bearing stars) can be ascribed to the smaller sample size for which $u$-band data are available, and to the intrinsically more scattered color distribution of disk-free stars on the $(u-r,~r)$ diagram, which prevents an accurate empirical determination of the reference color sequence for non-accreting sources.

The classification results obtained for individual objects from different accretion diagnostics show a substantial agreement. Namely, the \Eur{} classification is consistent with results from the \ErHa{}$^{CMD}$ (\ErHa{}$^{CCD}$) tracer in 67\% (82\%) of cases for which all indicators are available. 
The classifications produced by the \ErHa{}$^{CMD}$ and the \ErHa{}$^{CCD}$ tracers are highly correlated, overall yielding consistent results in 92\% of the cases.

\subsection{Variability of accretion tracers and final classification} \label{sec:accr_class}

The classification schemes described in Sects.~\ref{sec:E_ur}--\ref{sec:E_rHa} are based on the average photometric properties derived for each object during the VST/OmegaCAM observations. However, a more accurate depiction of how individual objects move across the accretion--dominated and photosphere--dominated regions of different color diagrams can be derived by monitoring the variability of the measured color excesses.

As detailed in \citet{Venuti2021}, the OmegaCAM survey comprised 17 distinct observing epochs, spread over 3.5 weeks. For each epoch, we measured instantaneous values of the color excesses, and then used the classifications extracted from the average photometry as a guide to assess the typical behavior of accreting, non-accreting, and potentially accreting stars around the accretion thresholds on the various diagrams. Common statistical trends could be identified among accreting and non-accreting objects across the various diagnostics. Namely, 90\% of stars classified as non-accreting were consistently found in the photosphere--dominated regions of the color diagrams in over 70\% of observing epochs. Conversely, 75\% of objects classified as accreting populated the accretion--dominated regions of the color diagrams in over 80\% of epochs. We then merged the single-epoch accretion classifications as follows:
\begin{itemize}
    \item for each individual color diagnostics, objects were flagged as accreting, non-accreting, or potential accretors based on the above statistics (i.e., accreting if displaying a color excess in $>$80\% of epochs, non-accreting if appearing as such in $>$70\% of epochs, and potential accretors otherwise);
    \item when all three indicators provided a coherent classification, the leading classification was retained as final accretion status for the corresponding object;
    \item if the three indicators provided discrepant classifications (some as accreting sources and some as non-accreting sources), the prevailing class was retained if supported by at least one indicator from each color diagnostics ($u-r$ and $r-H\alpha$), otherwise the object was labeled as a potential accretor;
    \item the resulting list of potential accretors was subsequently filtered, and a more definite accretion status was assigned when reasonable, by considering the prevalence of single-epoch measurements that would classify the object as accreting or non-accreting, and downgrading flags referred to color excess measurements with large uncertainties.
\end{itemize}

\begin{figure}
\centering
\includegraphics[width=0.47\textwidth]{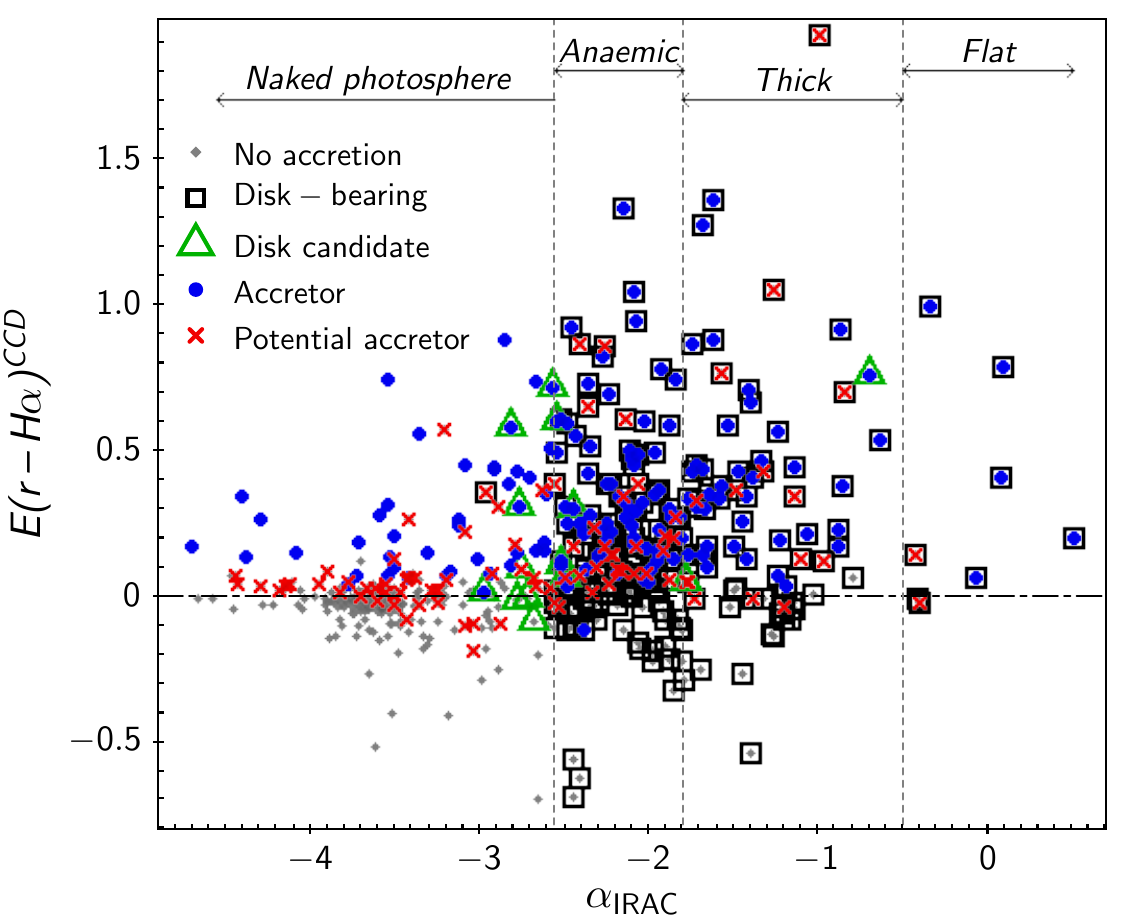}
\caption{Distribution of Lagoon Nebula YSOs according to their accretion properties (as traced via \ErHa$^{CCD}$, Sect.~\ref{sec:E_rHa}) vs. disk properties (as traced via $\alpha_{IRAC}$, Sect.~\ref{sec:disk_class}). Colors and symbols denote the final accretion/disk classifications derived for each object from our analysis: small gray dots are stars with no evidence of accretion from all available indicators; blue dots are accreting sources; red crosses are potential accretors; black squares surround disk--bearing sources; green triangles surround disk candidate sources. The disk class labels on top of the figure follow the $\alpha_{IRAC}$ scheme by \citet{Teixeira2012}.}
\label{fig:alphaIRAC_rHaExc}
\end{figure}

Our final aggregate classification, {reported in Table~\ref{tab:sample_properties}}, encompasses 343 accreting sources, 487 non-accreting sources, and 149 potential accretors. As in Sect.~\ref{sec:disk_class}, we followed a conservative approach to estimate the true statistical fraction of accreting stars ($f_{acc}$) in the Lagoon Nebula, accounting for classification uncertainties and missing data. This approach resulted in $f_{acc} = 0.38 \pm 0.10$ across the entire population, and $f_{acc} = 0.41 \pm 0.11$ when considering \textit{bona fide} members only. Accreting sources represent 54\% of targets that are classified as high-confidence disk-bearing stars, and 17\% of targets that are classified as high-confidence disk-free stars. {Nearly 86\% of high-confidence, disk-bearing sources classified as accreting were detected in accretion at all epochs during our monitoring campaign, whereas about one third of disk-free stars in the accreting group were detected in accretion only at certain epochs, indicating that these objects may be currently evolving past the disk accretion stage.} Non-accreting sources represent 26\% of our sample of high-confidence disk-bearing stars, and 68\% of our sample of high-confidence disk-free stars. A comparison between the final disk status and accretion status classifications across our sample of YSOs, with respect to $\alpha_{IRAC}$ and \ErHa$^{CCD}$, is illustrated in Fig.~\ref{fig:alphaIRAC_rHaExc}.

Disk-free stars flagged as accreting are concentrated predominantly (in $\sim$83\% of cases) at later spectral types (G/K/M), exceeding the statistical proportion of G-to-M types that is found across the entire population of disk-free stars. {While an intense chromospheric activity may, in some cases, affect the selection of individual objects as accretors based on the $H\alpha$ diagnostics, independent spectroscopic information available from the literature across this subset tends to support the reliability of our statistical classification. In particular, \citet{prisinzano2019} used the VLT/FLAMES spectra obtained for NGC~6530 members within the GES program to measure the full width at zero intensity (FWZI) of the H$\alpha$ line, which provides an accretion indicator where the impact of nebular contamination to the stellar spectra is minimized compared to other H$\alpha$ diagnostics that rely on the line core intensity \citep[e.g.,][]{bonito2020}. Among the 29 targets that represent the intersection between our high-confidence disk-free sample, our sample of accreting stars, and the sample of NGC~6530 members with FWZI measurement from \citet{prisinzano2019}, 23 (79\%) are located above the FWZI\,=\,4\,\AA{} threshold adopted in that paper to select candidate accretors, and 19 (66\%) have FWZI\,$>$\,5\,\AA, quoted by the authors as the empirical value above which known disk-bearing/accreting stars in their sample were distributed. Additional H$\alpha$ spectroscopic measurements from VLT/FLAMES spectra simultaneous with \emph{K2} Campaign~9 (program 097.C-0387(A), PI = S. Alencar) are available for a small subset (40 objects) of our target list, which includes only two stars classified as disk-free but accreting in our sample. Our photometric classification of sources as accreting, potential accretors, or non-accreting agrees with the derived spectroscopic information for all except five objects across this subset (88\%). Of the two disk-free sources with photometric indicators of accretion included in the subset, at least one (EPIC~248368907) shows clear spectroscopic evidence of accretion as well, with a measured H$\alpha$ width at 10\% intensity $\sim$600~km/s.} 

The nature of these objects with seemingly contrasting disk/accretion classifications could be understood, at least partly, in terms of more evolved circumstellar disks. As reviewed by \citet{espaillat2014}, young clusters and star forming regions as young as a few Myr are expected to comprise $\sim 3.5\%-7\%$ transition disk sources with large inner cavities that are no longer traceable in the near-IR. If applied to our sample, this would translate to $\sim 35-70$ transition disk candidates across the entire population, and $\sim 30-60$ across our subset of objects with IR data. A vast fraction of transition disk sources in young open clusters can exhibit accretion levels comparable to those of primordial disk sources \citep[e.g.,][]{sousa2019}. For about two thirds of the accreting, disk-free stars in our sample, only $J,H,K$ photometry was available to categorize their evolutionary status, which implies a putative transition disk around them could not have been detected. The remaining third had available \textit{Spitzer/IRAC} photometry that identified them as naked photospheres following \citeauthor{Teixeira2012}'s (\citeyear{Teixeira2012}) scheme. While the corresponding $\alpha_{IRAC}$ values are somewhat smaller than the typical anaemic disk indices extracted for transition disks in \textit{IRAC} bands, similar $\alpha_{IRAC}$ ranges have been tabulated by \citet{sousa2019} for about 20\% of their transition disk candidates.

\section{Disk accretion in the Lagoon Nebula}\label{sec:accretion_activity}

\subsection{Accretion rates and connection with $M_\star$} \label{sec:Mdot}

To convert the color excesses into estimates of the accretion luminosity $L_{acc}$, we followed the approaches of \citet{venuti2014} and \citet{Kalari2015}. Namely, we defined the flux excess due to accretion, $F_{exc}$, as
\begin{equation} \label{eq:Fexc}
    F_{exc} = F_{obs} - F_{phot},
\end{equation}
where $F_{obs}$ is the stellar flux calculated from the dereddened magnitude, and $F_{phot}$ is the contribution from photospheric and chromospheric activity. Assuming that the accretion luminosity only affects the $u$ and $H\alpha$ observations (not the $r$-band), the photospheric $u$ and $H\alpha$ magnitudes can be defined as $u_{phot} = u_{obs}$\,--\,\Eur{} and $H\alpha_{phot} = H\alpha_{obs}$\,+\,\ErHa{}. From Eq.~\ref{eq:Fexc}, we can then derive $F_{exc}^u$ and $F_{exc}^{H\alpha}$ as 
\begin{equation}
    F_{exc}^u = F_0^u \cdot 10^{-0.4 \cdot u_{obs}} \left(1-10^{+0.4 \cdot E(u-r)}\right)
\end{equation}
and
\begin{equation}
    F_{exc}^{H\alpha} = F_0^{H\alpha} \cdot 10^{-0.4 \cdot H\alpha_{obs}} \left(1-10^{-0.4 \cdot E(r-H\alpha)}\right),
\end{equation}
where $F_0^u$ ($1.2445 \times 10^{-6}$ erg/cm$^2$/s) and $F_0^{H\alpha}$ ($1.84 \times 10^{-7}$ erg/cm$^2$/s, corresponding to $H\alpha$-mag $\simeq$ 0.03;  \citealp{drew2014}) are the $u$-band and $H\alpha$-band integrated zero-point fluxes, respectively. The excess luminosity $L_{exc}$ is then calculated as $L_{exc} = F_{exc} \cdot 4\pi d^2$, where $d$ is the adopted distance to the cluster. To convert the filter--specific $L_{exc}$ into total accretion luminosity $L_{acc}$, we followed the prescription by \citet{venuti2014} for the $u$-band excess measurement, and the calibration relationships between $L_{exc}^{H\alpha}$ and $L_{acc}$ derived by \citet{alcala2017} for the T~Tauri regime and by \citet{fairlamb2017} for the Herbig (A/B) regime, after verifying that all A/B stars in our sample match the ranges in  mass and absolute magnitude presented respectively in \citet{vioque2022} and \citet{vioque2020} for confirmed and candidate Herbig Ae/Be stars. Finally, the mass accretion rate $\dot{M}_{acc}$ is estimated as
\begin{equation}
    \dot{M}_{acc} = \left(1 - \frac{R_\star}{R_{in}}\right)^{-1} \frac{L_{acc} R_\star}{G M_\star},
\end{equation}
where $G$ is the gravitational constant and $R_{in}$ is the infall radius for the accretion column, assumed to correspond to the truncation radius at $\sim$5\,$R_\star$ in the T~Tauri regime and $\sim$2.5\,$R_\star$ in the Herbig regime \citep{mendigutia2011}.

The $\dot{M}_{acc}$ values inferred from the three color diagnostics are largely consistent with each other, with the following median offsets:
\begin{itemize}
\item $\log{\dot{M}_{acc}^{(u-r)}} - \log{\dot{M}_{acc}^{(r-H\alpha)_{CCD}}} \simeq 0.1_{-0.4}^{+0.5}$
\item $\log{\dot{M}_{acc}^{(u-r)}} - \log{\dot{M}_{acc}^{(r-H\alpha)_{CMD}}} \simeq -0.1_{-0.3}^{+0.4}$
\item $\log{\dot{M}_{acc}^{(r-H\alpha)_{CCD}}} - \log{\dot{M}_{acc}^{(r-H\alpha)_{CMD}}} \simeq 0.04_{-0.05}^{+0.06}$
\end{itemize}

\begin{figure}
\centering
\includegraphics[width=0.47\textwidth]{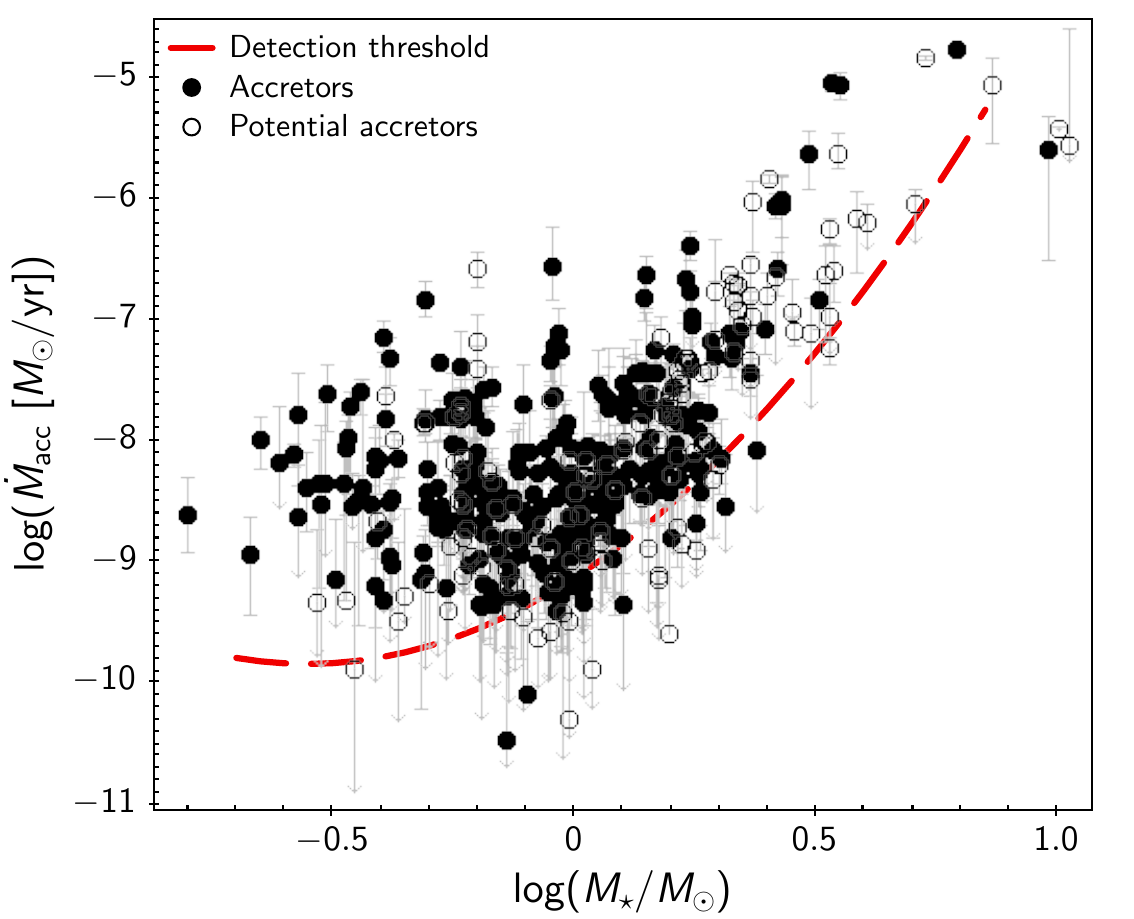}
\caption{Accretion rate distribution as a function of stellar mass for Lagoon Nebula stars classified as accreting (black dots) or potential accretors (empty circles). Variability bars mark the full range of $\dot{M}_{acc}$ values measured during the VST/OmegaCAM monitoring. Upper limits, marked as downward arrows, are assigned to objects that were located in the non-accreting regions of the various color excess diagrams at least in some epochs. The dashed red line traces a reference detection threshold, derived as a polynomial fit to the nominal $\dot{M}_{acc}$ values extracted for stars classified as non-accreting following the criteria discussed in Sect.~\ref{sec:accr_class}.}
\label{fig:M_Macc_all}
\end{figure}

A final, typical value of $\dot{M}_{acc}$ was assigned to each source by taking the median of the central values derived from each color diagnostics, while variability bars were derived by calculating the range between median and minimum/maximum values of the instantaneous $\dot{M}_{acc}$ measured along the entire monitored period with VST/OmegaCAM. These parameters are reported for all available targets in Table~\ref{tab:sample_properties}. Objects that fell below the accretion detection thresholds in color (see Sect.~\ref{sec:accretion_tracers}), at least at some epochs, were marked as upper limits with no defined lower variability bar. The resulting $\dot{M}_{acc}$ distribution as a function of $M_\star$ is illustrated in Fig.~\ref{fig:M_Macc_all}.

To extract any mass--dependent trends in $\dot{M}_{acc}$, we measured the median logarithmic $\dot{M}_{acc}$ in {0.2}~dex--wide logarithmic mass bins, using a {0.1}~dex--wide moving window. This analysis was only conducted in the mass range $-0.7 < \log{M_\star} < {0.8}$, to ensure the presence of at least {four} accreting sources in each bin, and revealed distinct $M_\star - \dot{M}_{acc}$ regimes: a substantially flat trend for $\log{M_\star} \mathrm{<0}$; a monotonically increasing trend in $\log{\dot{M}_{acc}}$ for $\log{M_\star}$ between ${0}$ and {0.6}; and a {potential flattening in the $M_\star - \dot{M}_{acc}$ trend that begins to emerge at $\log{M_\star} \geq 0.6$}. We then followed the approach discussed in \citet{vioque2022} to refine our search for the transition values of $\log{M_\star}$. Namely, we let the test value of logarithmic transition mass vary across the entire range covered by two consecutive regimes, as identified earlier (i.e., from $-0.7$ to {0.6} in the first case, and from {0} to {0.8} in the second case). We used each test value to split the corresponding sample of objects into two mass groups, and we applied a linear regression routine to calculate the best--fitting slope for both groups separately. We then estimated the significance of the slope difference between the two groups, weighted by the uncertainty on the derived slopes, and we built a distribution of this significance parameter as a function of the assumed transition mass. Finally, we defined our best estimate of logarithmic transition mass as the value that maximizes the significance of the difference in slope between the two neighboring mass groups. {This procedure allowed us to extract a robust estimate of the first transition point at $\log{M_\star} \simeq {0.01}$}, with a best--fitting slope of $\log{\dot{M}_{acc}} = \left(4.5\pm0.3\right)\cdot \log{M_\star} + k$ to describe the $M_\star-\dot{M}_{acc}$ relationship in the intermediate mass regime, and a typical value of $\log{\dot{M}_{acc}}\simeq -8.6$ for the flat distribution of accretion rates at masses $\log{M_\star} \leq 0.01$. 

{Given the small number of objects at the high mass end of our distribution, we could only obtain a rough lower limit of $\log{M_\star} \simeq {0.61}$ for the second potential transition point. We note that this lower limit (at $M_\star \sim 4.07\,M_\odot$)} is consistent with the better defined mass value at which a break in the $M_\star-\dot{M}_{acc}$ relationship for Herbig AeBe stars has been reported by \citet[][$M_\star = 3.98^{+1.37}_{-0.94}\,M_\odot$]{wichittanakom2020} and \citet[][$M_\star = 3.87^{+0.38}_{-0.96}\,M_\odot$]{vioque2022}. A positive correlation trend between $\log{\dot{M}_{acc}}$ and $\log{M_\star}$ was reported by the same authors for the higher--mass component of the Herbig AeBe population, albeit shallower than the one found among less massive Herbig AeBe stars. Such a trend cannot be decisively extracted from our data due to the dearth of objects at masses higher than $\log{M_\star} \sim 0.5$ in our sample. On the other hand, the slope extracted in the logarithmic mass range $\simeq 0-{0.6}$ is at least qualitatively consistent with the trend traced for lower--mass Herbig AeBe stars in \citet{vioque2022}, {and it overlaps with the slope intervals for the lower-mass group estimated by \citet{wichittanakom2020}}. The overall flat trend observed at the lower--mass end of our distribution in Fig.~\ref{fig:M_Macc_all} ($-0.5 \leq \log{M_\star} \leq 0$) qualitatively matches analogous trends observed over the same mass range in different star--forming regions \citep[e.g.,][]{venuti2014}, although the median $\log{\dot{M}_{acc}}$ value extracted here appears somewhat lower than those reported for other young stellar populations. This feature, coupled with the lower frequency of some irregular disk--driven variables (bursting, stochastic, aperiodic dippers) that are identified in the Lagoon \textit{K2} sample \citep{Venuti2021} compared to other young clusters surveyed from space \citep{cody2014,cody2018,cody2022}, may indicate more moderate accretion dynamics in the Lagoon Nebula cluster. 

When we analyze the typical accretion properties detected for specific variable groups with respect to those measured across the entire population as a function of mass, we do observe some correlation with the light curve morphology class. The one burster star with mass estimate in our sample is located in the top quartile of the $\dot{M}_{acc}$ distribution for the corresponding mass bin. Stochastic variables have typical accretion rates higher than the median measured across the entire population in 80\% of the mass bins where they are represented, while the same is true for flat-line variables in $\sim$60\% of mass bins where they appear; for aperiodic dippers in $\sim$55\% of mass bins; for periodic and multi-periodic variables and quasi-periodic dippers in $\sim$40\%--45\% of mass bins; and for quasi-periodic symmetric variables in 25\% of mass bins. 

\subsection{Wavelength--dependent trends in photometric light curves for accreting stars}\label{sec:timing_lambda}

In the magnetospheric accretion framework, hot spots at the stellar surface are predicted to be structured in density, with a more compact ($\lesssim$1\% area), high-density region emitting predominantly at shorter wavelengths, and a more extended ($\gtrsim$10--20\%), low-density region emitting predominantly at longer wavelengths \citep[e.g.,][]{romanova2015}. This gradient in the radial density profile is expected to translate to a wavelength-dependent apparent size for the hot spot and to a slightly asynchronous nature in the spot-induced variability patterns at different filters. Recently, a first direct observational confirmation of the latter prediction has been obtained for the young accreting star GM~Aur by \citet{espaillat2021}, who monitored the star from the $u$-band to the $i$-band and detected a significant delay in the appearance of luminosity peaks at optical wavelengths with respect to their timing in the near-UV.

Unfortunately, the sparser cadence of our ground-based monitoring data does not allow us to perform a precise peak timing analysis. However, we did conduct a statistical exploration of wavelength--dependent trends in the light curve morphology of our accreting targets by taking as reference the \textit{K2} time series. More specifically, after rejecting $>$5\,$\sigma$--discrepant points and removing any underlying systematic trends, all \textit{K2} light curve points between 1\textsuperscript{st}--99\textsuperscript{th} percentiles in flux were selected, and a brightness phase was defined as 
\begin{equation}
    \phi_i^{flux} = \frac{{f_i} - {f_{min}}}{{f_{max}} - {f_{min}}},
\end{equation}
where $f_i$ is the normalized flux at epoch $i$, and $f_{min}$ and $f_{max}$ correspond to the percentile levels defined earlier. We then matched each VST/OmegaCAM observing epoch with the closest \textit{K2} observation, and defined a corresponding value of expected $\phi^{flux}$ as the median $\phi_i^{flux}$ within a range of five \textit{K2} epochs around the best-matching date. 

To assess the overall agreement between the photometric variations traced at $u,g,r,i$ wavelengths and the \textit{K2} light curves, we applied a Spearman's rank correlation test, as implemented in the Python module \texttt{scipy.stats}, to the array of (mag,~$\phi^{flux}$) values obtained for each band. In case of a perfect correspondence between the \textit{K2} light curve pattern and the magnitude trends in a given VST/OmegaCAM filter, we would expect to observe a definite anticorrelation trend between the two quantities in the array  (i.e., lower values of magnitude corresponding to higher values of $\phi^{flux}$), leading to a negative rank correlation coefficient $\rho$ that approaches unity. Conversely, OmegaCAM and \textit{K2} light curves in opposition of phase would lead to positive $\rho$ values, because increases in magnitude in the ground-based time series would be matched with large $\phi^{flux}$ values (extracted from the \textit{K2} time series).

\begin{figure}
\centering
\includegraphics[width=0.5\textwidth]{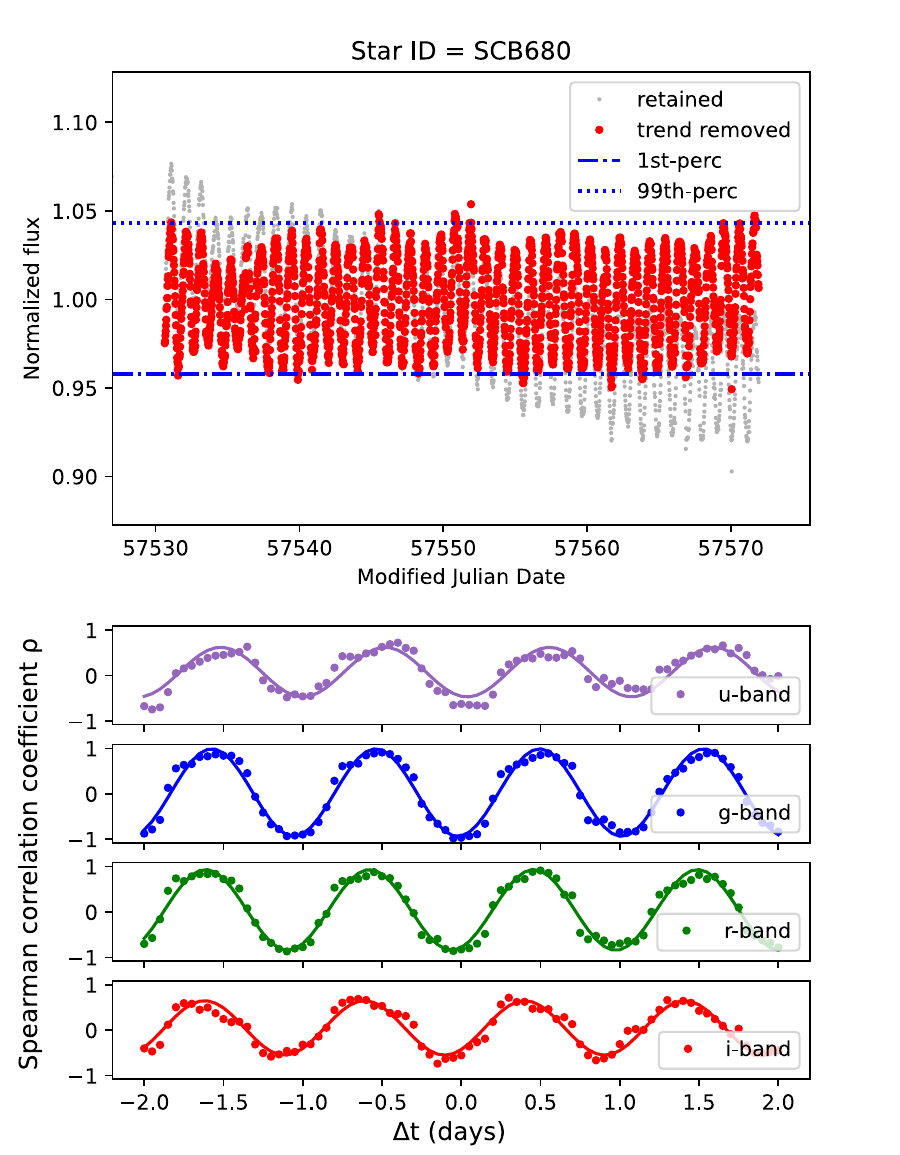}
\caption{Example of a periodic \textit{K2} light curve for a young, accreting star in the Lagoon Nebula (top), and results of the correlation analysis between the \textit{K2} time series and each of the $u,g,r,i$ light curves as a function of the  hypothetical time lag between the respective variation patterns (bottom), as described in the text. The overlaid curves in the bottom panel represent sinusoidal fits to the measured ($\Delta t$, $\rho$) trends.}
\label{fig:rho_analysis_example}
\end{figure}

To probe whether the agreement between \textit{K2} and each of the individual $u,g,r,i$ light curves would be improved by assuming a time shift between the two time series, we repeated the analysis described above by imposing a lag $\Delta t$ when matching the VST and \textit{K2} epochs. In our exploration, we let $\Delta t$ vary between $-2$~days and $+2$~days, and at each step we assigned a best-matching \textit{K2} epoch ($t_{K2}$) to each VST epoch ($t_{VST}$) as the one that minimized $\lvert t_{VST}-(t_{K2}-\Delta t) \rvert$. For each object and filter, we then examined the resulting values of $\rho$ as a function of $\Delta t$. An illustration of this procedure for the accreting young star SCB680 is shown in Fig.~\ref{fig:rho_analysis_example}.

For regular variables, an underlying periodic pattern in the ($\Delta t$, $\rho$) trend is expected, corresponding to the rotation rate $P_{rot}$ of the star. Small deviations from that periodicity may reveal distinctive trends in the appearance of variability features as a function of wavelength. To extract any such trends from our light curves, we conducted a sinusoidal fit on the ($\Delta t$, $\rho$) distribution filter by filter, using the \texttt{scipy} function \texttt{optimize.curve\_fit}, and defining our model function as
\begin{equation}
    \rho = a\cdot \sin(b\cdot \Delta t - c) + d.
\end{equation}
We restricted this exploratory analysis to accreting YSOs in the Lagoon Nebula with light curves classified as periodic or quasi-periodic, and assigned initial values for the fit parameters as $a_0 = 0.5\cdot (\rho_{max} - \rho_{min})$, $b_0 = 2\pi/P_{rot}$, $c_0 = 0$, and $d_0 = 0.5\cdot (\rho_{max} + \rho_{min})$. To interpret the results, we then focused on the best-fit parameters $b$ and $c$, which are related to the periodicity $P$ and time shift $\tau$ of the sinusoid as $P = 2\pi/b$ and $\tau = -c/b$. In this formulation, negative values of $\tau$ indicate a delay in the sinusoid.

A successful sinusoidal fit with an optimal solution was achieved for 27/40 accreting stars with well-behaved, regular light curves in our sample. Some interesting trends emerged from these results. In 21/27 cases, a definite trend in the measured $P$ as a function of wavelength could be identified: in 57\% of the cases, the measured $P$ overall increases from the $u$-band to the $i$-band (typical change $\sim$18\% of the $u$-band period), while in the remaining 43\% of cases, $P$ is found to decrease from $u$ to $i$ (with a smaller typical change of $\sim$6\% in $P$). {For comparison, the maximum uncertainties on $P$, as measured from the \emph{K2} light curves, amount to 4\%--8\% for the least periodic cases in our subset, and for a case such as the one illustrated in Fig.~\ref{fig:rho_analysis_example} the estimated uncertainty on $P$ is $<$1\%.} No obvious connection was observed between the specific trend in $P$ and stellar parameters like $M_\star$ and SpT; 
however, stars in this subset that were consistently observed as accreting at all VST/OmegaCAM epochs were found to exhibit preferentially an increase in $P$ from $u$ to $i$ (77\% of the cases), whereas 71\% of stars in the subset that only appeared as accreting at some epochs were found to exhibit $P$ that decreases from $u$ to $i$.

While a detailed modeling of these behaviors is beyond the scope of this work, we note that drifts in the measured period as a function of wavelength may reflect slightly different intrinsic timescales for the features that dominate the emission detected in separate filters. As discussed in \citet[][see in particular their Fig.~4]{espaillat2021}, the differential rotation between the stellar magnetosphere and the inner disk may induce a dragging effect on the accretion funnels, triggering changes in both the shape and location of the hot spots. As a consequence, the hot-spot modulation pattern would be slightly out of sync with respect to the rotation rate of the star. Because the structure of the densest hot-spot regions are particularly affected by these dynamics, non-stationary effects in variability would be observed more predominantly in bluer filters, leading to potential differences in the characteristic timescales extracted from different wavelength regimes.

Regarding $\tau$, a time shift between the $u$-band and the $i$-band was detected to a significance of at least $\sim$2\,$\sigma$ for 18/27 stars in our subset. For the vast majority (16, i.e., $\sim$90\%) of them, a positive $\Delta \tau$ between $u$ and $i$ was measured, indicating that the $u$-band variability pattern is ahead with respect to the $i$-band variability pattern. The median time difference between $u$ and $i$ across the 16 targets amounts to $\sim$7\% of the corresponding rotation period measured, for each star, from the \textit{K2} light curve. However, these statistics are highly correlated with the specific $P$ vs. wavelength trends discussed above: among stars with $P$ decreasing from the $u$-band to the $i$-band, we measured a typical $\Delta \tau/P_{rot} = 0.34 \pm 0.06$, while among stars with $P$ increasing from $u$ to $i$, we found $\Delta \tau/P_{rot} = 0.06 \pm 0.04$.

\subsection{{Variability and time dependence of accretion}}\label{sec:Macc_var}

For 275 sources across our sample, we have recorded detectable accretion levels at all epochs during the OmegaCAM monitoring. The median $\dot{M}_{acc}$ variability amplitude (and relative dispersion) measured across this subset of objects amounts to $0.5^{+0.3}_{-0.2}$~dex, consistent with results reported earlier for various young stellar populations \citep[e.g.,][]{venuti2014,costigan2014}. Slightly different variability amplitudes above ($\Delta{\dot{M}_{+}}$) and below ($\Delta{\dot{M}_{-}}$) the median $\dot{M}_{acc}$ were statistically measured across the sample ($0.21^{+0.11}_{-0.09}$~dex and $0.28^{+0.28}_{-0.14}$~dex, respectively), but the two values overlap well within the scatter. No significant dependence of the $\dot{M}_{acc}$ variability amplitude on $M_\star$ was observed.

Some marginal trends in $\dot{M}_{acc}$ variability can be extracted from the comparison with the light curve morphological classification for sources with \textit{K2} data. Flat-line variables tend to exhibit the largest amplitudes of $\dot{M}_{acc}$ variability over rotational timescales ($\Delta{\dot{M}}\sim0.6$~dex), followed by quasi-periodic symmetric and aperiodic dipper variables ($\Delta{\dot{M}}\sim0.5$~dex), periodic and stochastic variables ($\Delta{\dot{M}}\sim0.4$~dex), quasi-periodic dippers ($\Delta{\dot{M}}\sim0.3$~dex), and our only burster variable with $\dot{M}_{acc}$ estimate ($\Delta{\dot{M}}\sim0.2$~dex). Larger $\Delta{\dot{M}}$ would indicate a stronger dependence of the observed accretion rate on the instantaneous viewing geometry, or more intermittent accretion levels on the inner disk dynamical timescales. Smaller $\Delta{\dot{M}}$, on the other hand, would imply a nearly constant visibility of surface accretion features or a homogeneous distribution of accretion streams, coupled with accretion dynamics able to sustain similar $\dot{M}_{acc}$ levels over several rotational cycles.

To assess any longer-term $\dot{M}_{acc}$ variability trends, we compared our results with those obtained by \citet{Kalari2015}, who measured accretion rates for low-mass stars in the Lagoon Nebula using observations taken, four years earlier, with the same instrument (OmegaCAM). We cross-correlated our catalog with theirs, and could retrieve 100 sources common to both samples (within a 1$''$ matching radius) and with detected $\dot{M}_{acc}$ in both studies. Across this subset of objects, the typical difference between the single-epoch $\dot{M}_{acc}$ measurement reported by \citet{Kalari2015} and the median value of $\dot{M}_{acc}$ reported here amounts to $\Delta (\log{\dot{M}_{acc}}) = 0.55 \pm 0.35$ in the logarithmic mass range from $-0.4$ to 0.4. This value is well consistent with the combined $\log{\dot{M}_{acc}}$ uncertainties and variability bar for a typical star common to this and \citeauthor{Kalari2015}'s (\citeyear{Kalari2015}) study ($\sim$0.75~dex).

No specific trends in years-long $\Delta (\log{\dot{M}_{acc}})$ with light curve behavior (modulated vs. erratic vs. dipping) emerged among stars with long-term variability information and {\em K2} light curves (34 objects).
However, we note that only three YSOs in this subset ($\sim$9\%) exhibit a years-long difference in $\dot{M}_{acc}$ that stands over three times above the typical population value in units of the associated dispersion, and all three of them (EPIC IDs 224321494, 224324884, 224321735) belong to the erratic variable group.

\begin{figure}
\centering
\includegraphics[width=0.47\textwidth]{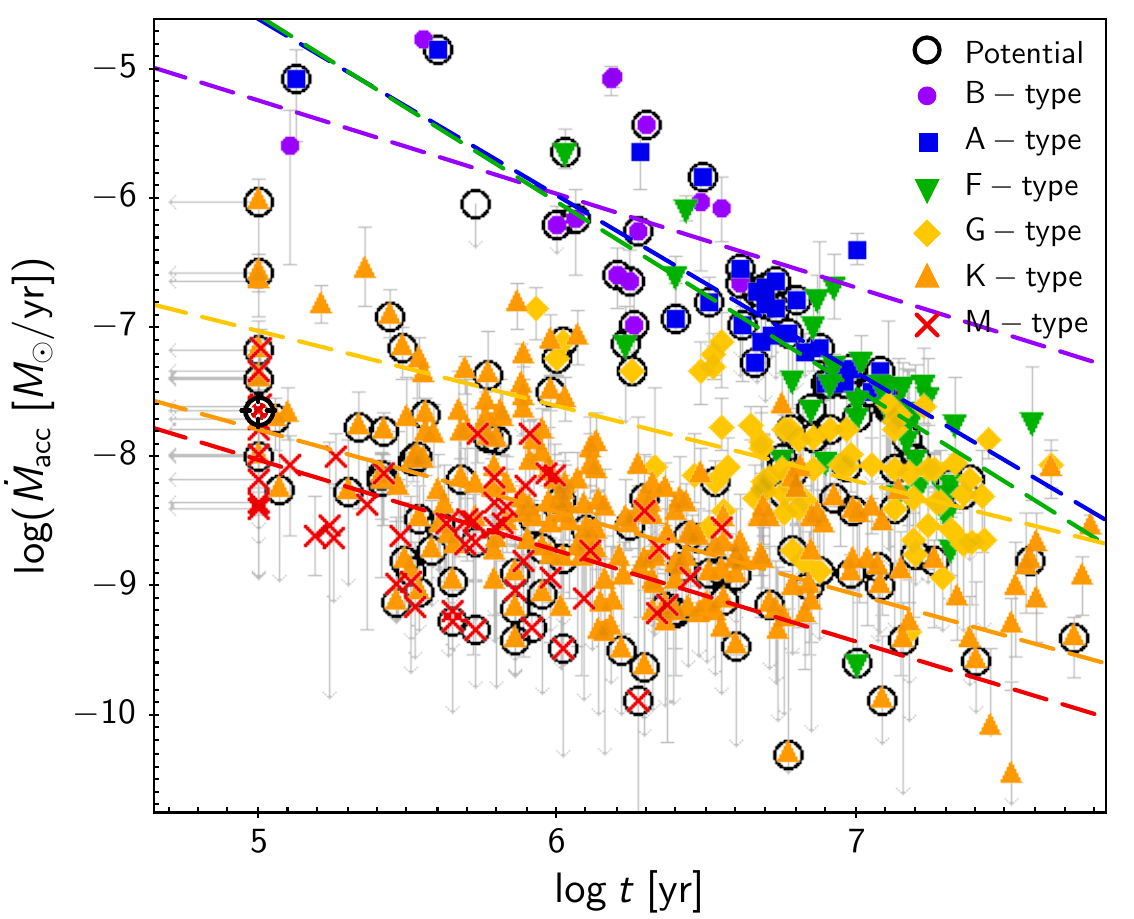}
\caption{{Correlation trend between the estimated ages and $\dot{M}_{acc}$ values for accreting Lagoon Nebula members as a function of spectral class (colors and symbols as labeled in the legend). Stars classified as potential accretors are encircled in black. The dashed lines mark the linear trend corresponding to the best-fit slope and intercept derived using the Theil-Sen estimator, with colors matching the corresponding spectral class.}}
\label{fig:age_Macc_SpT}
\end{figure}

{Finally, we examined the connection between the typical $\dot{M}_{acc}$ values measured for sources across the Lagoon Nebula and their estimated ages, as derived in Sect.~\ref{sec:HR_diagram}. While individual ages extracted for pre-main sequence YSOs on the HR diagram are notoriously uncertain \citep[e.g.,][]{soderblom2014}, the fundamental stellar and accretion properties that we determined within our sample do support an evolutionary scenario reflected in the population spread across the isochrone grid on Fig.~\ref{fig:Lagoon_HR_diagram_MIST_nonrot_solar}. Namely, when dividing our accreting sources into spectral classes, a statistically significant anticorrelation trend emerges between individual $\dot{M}_{acc}$ and isochronal ages in each group {from A to M stars, with $p$-values $<$ 0.001 resulting from a two-sided Spearman's rank-order correlation test and Kendall's rank correlation coefficient, as verified by performing a permutation test. These anticorrelation trends are illustrated in Fig.~\ref{fig:age_Macc_SpT}, where the linear fits overlaid on each SpT group were derived by using the Theil-Sen estimator to obtain an unbiased approximation of the true slope.} This association would suggest the presence of an intrinsic age spread among Lagoon Nebula YSOs, with accretion decreasing as stars become older \citep[e.g.,][]{beccari2010}. {The lower agreement between the datapoint distribution and the simple linear regression result for B-type stars reflects the smaller sample size and comparatively larger scatter, which translates to a more uncertain anticorrelation trend with a $p$-value $\sim$ 0.15.}}

\section{Discussion}\label{sec:discussion}

\subsection{Mass dependence of disk and accretion properties}\label{sec:facc_fdisk_mass}

\begin{figure*}
\centering
\includegraphics[width=\textwidth]{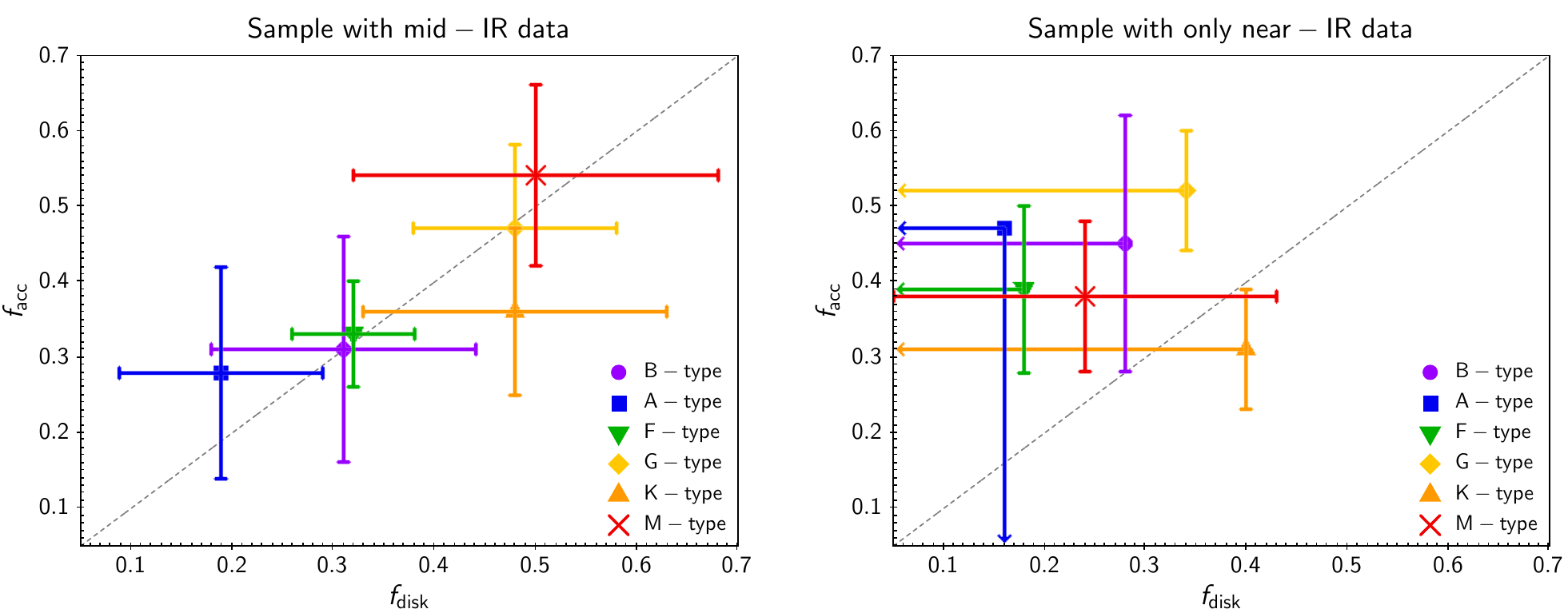}
\caption{Comparison between the statistical, SpT--dependent fractions of disk--bearing and accreting stars measured across the Lagoon Nebula population, {as estimated from the subset of objects with disk classifications based on \textit{Spitzer}/IRAC data (left) and on the subset for which only near-IR $J,H,K$ data were available (right)}. O-type stars do not appear on the diagrams because no definite estimate of $f_{disk}$ could be extracted for them. {Arrows denote upper limits, and} the equality line is dotted in gray.}
\label{fig:SpT_fdisk_facc}
\end{figure*}

To investigate any connection between disk activity and stellar mass across the Lagoon Nebula, we used the statistical approach presented in Sect.~\ref{sec:disk_class} to extract SpT--dependent estimates of $f_{disk}$ and $f_{acc}$ across our sample. Only stars with both disk and accretion indicators available (814) were used to define the observed sample size and measured fractions. To account for the mass--dependent completeness levels shown in Fig.~\ref{fig:mass_distribution}, we used the mass distribution of young stars with measured parameters in each spectral class to estimate the number of missing objects. We then introduced this estimate as a correction to the true population size, which was used to simulate the intrinsic $f_{disk}$ and $f_{acc}$ that are statistically compatible with the observed fractions of disk-bearing and accreting objects as a function of SpT. {Moreover, to explore the timescales of disk clearing with respect to the duration of the accretion phase, we conducted this analysis separately for the subset of stars in our sample for which mid-IR data from \textit{Spitzer}/IRAC could be used to determine the disk status (496) and for those for which only near-IR data (2MASS) were available (318).}

The results of this statistical exploration are shown in Fig.~\ref{fig:SpT_fdisk_facc}. Although individual points overlap, due to the combined uncertainties related to sample incompleteness and non-definite classifications (i.e., disk candidates, potential accretors), some {mass--dependent} trends emerge from this picture. Along the horizontal axis {on the left panel, which probes the prevalence of disks detectable at mid-IR wavelengths}, lower-mass stars (G/K/M spectral types) tend to exhibit higher values of $f_{disk}$ than more massive stars. The typical difference in the central $f_{disk}$ value estimated for G/K/M stars with respect to B/F stars amounts to $\sim${17}\%, or over {half} the estimate extracted for B/F stars. In other words, disks around G/K/M stars appear to be about {1.5} times more frequent than around B/F stars. {The statistical estimates of $f_{acc}-f_{disk}$ ratios are distributed around the equality line on this diagram, indicating a substantial agreement in the evolutionary dynamics probed by $\alpha_{IRAC}$ and accretion diagnostics.}

{A different picture appears on the right panel of Fig.~\ref{fig:SpT_fdisk_facc}, which illustrates the statistical information extracted from stars for which only $J,H,K$ photometry could be used to assess the disk status. Compared to the mid-IR sample depicted on the left, a clear drop in the estimated $f_{disk}$ ranges can be observed across all mass groups, which suggests that the process of inside-out disk clearing has already begun at the Lagoon Nebula age. However, the statistical $f_{acc}$ ranges estimated from the near-IR subset are consistent with those determined from the mid-IR subset across most SpT groups, resulting in an overall shift of the $f_{acc}-f_{disk}$ ratios, which largely lie above the equality line on this diagram.}
Such a discrepancy between the measured $f_{acc}$ and $f_{disk}$ may {reflect} distinct evolutionary timescales for the gas and dust contents of the {inner regions of} protoplanetary disks. Indeed, recent numerical studies \citep[e.g.,][]{appelgren2020} have shown that, following an initial phase of slow dust drifting at $t<1$~Myr, the dust within the disk can be drained quickly even as the gas accretion rate remains approximately constant. {Several factors can contribute to setting the duration of the dust disk clearing phase, including the disk mass} (with more massive disks being more common around higher-mass stars; \citealp[e.g.,][]{rilinger2023}), and the photoevaporation induced by the stellar X-ray luminosity \citep[e.g.,][]{kimura2016}, which is typically more intense for higher-mass YSOs compared to lower-mass objects \citep[e.g.,][]{jardine2006}. {Although no conclusive relative ordering of $f_{acc}-f_{disk}$ ratios as a function of SpT can be extracted from the right panel of Fig.~\ref{fig:SpT_fdisk_facc}, a faster disk evolution for higher-mass stars than for lower-mass stars in the Lagoon Nebula is suggested by the difference in steepness of the respective age--dependent trends in Fig.~\ref{fig:age_Macc_SpT}.}

The measurement of higher $f_{disk}$ among lower-mass stars is overall consistent with the results of earlier statistical surveys of disk fractions that employed similar photometric criteria for disk classification. In particular, \citet{ribas2015} analyzed a sample of $\sim$1400 young stars from 22 nearby star-forming regions and found that disk evolution occurs more rapidly around higher-mass stars ($M_\star > 2 M_\odot$) than around lower-mass stars. This effect could already be discerned at very young ages ($\sim$1--3 Myr): indeed, low-mass YSO populations falling into that age bin in \citeauthor{ribas2015}'s (\citeyear{ribas2015}) sample were found to exhibit a fraction of protoplanetary disks that amounted to over 1.6 times the $f_{disk}$ measured around higher-mass stars in the same regions. We note that, while this trend provides a qualitative match to our findings, the absolute values of $f_{disk}$ extracted by \citet{ribas2015} on nearby stellar associations and star-forming regions are significantly higher than the disk fractions derived here. Our estimate in Sect.~\ref{sec:disk_class} is, however, consistent with the prevalence of infrared excess sources among the total population of probable Lagoon Nebula members reported by \citet{Povich2013} from the MYStIX project \citep{feigelson2013}. This would suggest that \citeauthor{ribas2015}'s (\citeyear{ribas2015}) target regions may be subject to very different environmental feedback on disk evolution than the Lagoon Nebula as a whole, although evidence for a possible gradient in conditions across the latter complex has also been reported \citep[e.g.,][]{richert2018}.

In a complementary approach, \citet{vandermarel2021} investigated the evolution of protoplanetary disks as a function of the host star mass by looking at the occurrence of disk substructures from ALMA data. In their survey, also focused on young, nearby stellar associations, the authors observed that the percentage of structured circumstellar disks (defined as featuring gaps, cavities, and rings, as opposed to compact disks that are largely non-structured) increases with stellar mass. Below $M_\star = 1 M_\odot$, compact disks were found to represent 75--90\% of all observed disks, while for more massive stars structured disks were observed in 50--60\% of the cases. Because structured disks tend to be more extended than compact disk and the dust radius is comparatively found at larger distances from the stars, such disks would move quickly past the detection threshold in the near- to mid-IR diagnostics that are used here to identify disk--bearing sources. This picture is also qualitatively consistent with the statistically higher detection rate of disks around lower-mass stars than around high-mass stars that we derived for the Lagoon Nebula.

\subsection{Environmental feedback on disk evolution}\label{sec:facc_fdisk_location}

\begin{figure}
\centering
\includegraphics[width=0.47\textwidth]{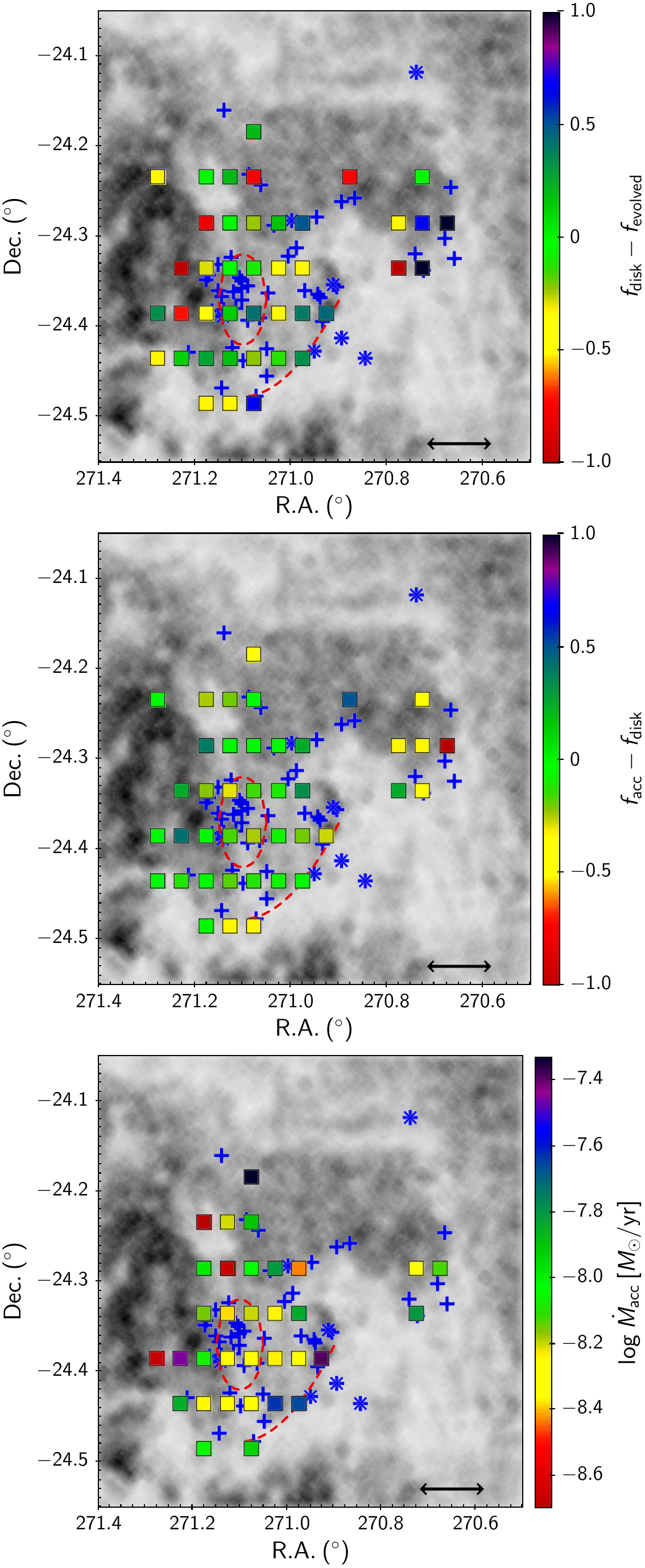}
\caption{Spatial variations of the measured $f_{disk}$, $f_{acc}$, {and $\dot{M}_{acc}$} across the Lagoon Nebula region (North up, East left). On all panels, the background density map illustrates the distribution of point sources extracted from the VST/OmegaCAM images. The {top} panel shows local differences in proportion of disk--bearing ($f_{disk}$) vs. disk--free ($f_{evolved}$) sources when the field is divided into a grid of $0.05^\circ\times0.05^\circ$ bins. Each bin is color--coded according to the measured $f_{disk}-f_{evolved}$, following the color scheme of the side axis. The {middle} panel shows the same field characterization where each bin is colored according to the local difference between $f_{acc}$ and $f_{disk}$. {The bottom panel illustrates local variations on the average measured $\log{\dot{M}_{acc}}$.} On all diagrams, O-type and B-type members are identified as {blue} asterisks and crosses, respectively. The double arrow in the bottom-right corner marks a reference distance of 3~pc at the location of the Lagoon Nebula. The two {red dashed} contours indicate the spatial concentrations of the youngest Lagoon Nebula YSOs identified by \citet{prisinzano2019}.\looseness=-1}
\label{fig:fdisk_facc_radec_map}
\end{figure}

To explore any local variations in the proportion of disk--bearing and accreting stars, we divided the field of view into $0.05^\circ$--wide (R.A., Dec.)\ bins on each side, and within each bin we counted the fraction of objects classified as either disk--bearing or disk--free and either accreting or non--accreting. {To minimize any selection biases on disk-bearing sources, we limited this analysis to the sample of objects with mid-IR data available for disk classification.} We then retained all bins with three or more projected members to build a map of the frequency of disk activity across the Lagoon Nebula.

Our results are illustrated in Fig.~\ref{fig:fdisk_facc_radec_map}. In the {top} panel, {a non-uniform spatial pattern} appears in the local fraction of disk--bearing stars with respect to disk--free sources ($f_{disk}-f_{evolved}$). {At the heart of NGC~6530}, within a couple of pc around {the central coordinates of the O/B population at} (R.A.,~Dec.) $\sim$ ($271.01, -24.35$), {disk-free or evolved disk sources appear to be statistically as numerous, if not more abundant, with respect to more primordial disks that can be detected at 1--12~$\mu$m wavelengths.} The balance between disk--bearing and accreting stars, instead, appears {to be more constant} across the same region, with a measured $f_{acc}$ similar to the local $f_{disk}$ ({middle} panel of Fig.~\ref{fig:fdisk_facc_radec_map}). A {potential} overabundance of disk--bearing stars with respect to more evolved stars {may emerge} along the south-west edge of the cluster, between (R.A.,~Dec.) $\sim$ ($271.0, -24.45$) and (R.A.,~Dec.) $\sim$ ($270.9, -24.35$), {and around a more detached agglomeration of members at (R.A.,~Dec.) $\sim$ ($270.7, -24.28$). The latter location appears to be characterized by} a prevalence of disk--bearing stars over accreting stars, {as suggested by the color scale on the middle panel of Fig.~\ref{fig:fdisk_facc_radec_map}}.

These trends seem to correlate, {at least qualitatively}, with the distinctive spatial distributions of accreting/disk--bearing vs. non-accreting/disk--free Lagoon Nebula members mapped in \citet{prisinzano2019}. Those authors noted two separate locations around which very young, accreting/disk--bearing members are preferentially concentrated: the first around (R.A.,~Dec.) $\sim$ ($271.1, -24.37$), close to the location with the highest concentration of massive stars on Fig.~\ref{fig:fdisk_facc_radec_map}, and the second following an elongated rim between (R.A.,~Dec.) $\sim$ ($271.1, -24.47$) and (R.A.,~Dec.) $\sim$ ($270.9, -24.35$), which overlaps with the {south-west cluster edge} described above. The approximate boundary of the first overdensity region from \citet{prisinzano2019}, as well as the lower envelope for the second, elongated locus, are drawn on both panels of Fig.~\ref{fig:fdisk_facc_radec_map}. \citeauthor{prisinzano2019} noted that, while non--accreting/disk--free objects are found in a more widespread distribution around the center of the NGC~6530 cluster (and hence overlap with the first concentration of accreting/disk--bearing members), very few evolved members are found near the elongated rim. They suggested that these spatial properties are indicative of a sequential star formation history triggered by two distinct ionizing fronts, one at the center of the NGC~6530 cluster, and the other traced by the south-west rim.

A similar spatial analysis of accretion intensity, {illustrated in the bottom panel of Fig.~\ref{fig:fdisk_facc_radec_map},} also revealed some notable trends. Within a radius of $\sim$1.2~pc around (R.A.,~Dec.) $\sim$ ($271.1, -24.38$), accreting cluster members typically display moderate $\dot{M}_{acc}$ levels, of order $\sim 5\times 10^{-9} M_\odot/yr$. Higher typical $\dot{M}_{acc}$ levels ($\sim 1\times 10^{-8} M_\odot/yr$) are instead measured outside of this region and within a radius of $\sim 3.5$~pc from the cluster center, with a peak $\dot{M}_{acc} \sim 3\times 10^{-8} M_\odot/yr$ around the south-west location at (R.A.,~Dec.) $\sim$ ($270.9, -24.4$). Comparable accretion rates $\dot{M}_{acc} \sim 7-15\times 10^{-9} M_\odot/yr$ are also found around the {western concentration of members at (R.A.,~Dec.) $\sim$ ($270.7, -24.3$)}. Taken together, the lower disk fractions and accretion rates that are observed around the core of NGC~6530 with respect to its outer shell suggest that the intense radiation field produced by massive O/B stars at the heart of the region {may} impact the pattern and timescales of disk evolution. The combined effects of externally--induced disk clearing and internal photoevaporation may lead to shortened disk lifetimes \citep[e.g.,][]{winter2022}, as traced from dust emission diagnostics. {Indeed, observational studies of other massive star forming regions such as NGC~2244 \citep{balog2007}, the Eagle Nebula \citep{guarcello2010}, and Cygnus~OB2 \citep{guarcello2023} have revealed a definite drop in the measured disk fractions within a radius $\sim0.5-1$~pc from the location of the highest-mass stars.} At the same time, gas accretion can still continue at a detectable rate across the developing disk cavity \citep[e.g.,][]{garate2021}, possibly explaining the lack of a drop {of similar magnitude} in $f_{acc}$ on the {middle} panel of Fig.~\ref{fig:fdisk_facc_radec_map} that would produce a color pattern matching the {top} diagram. An increased proportion of disk--bearing sources along the south-west cluster boundary, in spite of it being similarly aged as the NGC~6530 core \citep{prisinzano2019}, would then denote a reduced impact of external conditions on the local disk evolution pattern, with a potential restoration of the typical mutual ordering between accretion and disk dispersal timescales (with $f_{acc}$ decreasing more quickly than $f_{disk}$; e.g., \citealp{fedele2010,venuti2018}). 

{In order to test this scenario, we followed the prescriptions of \citet{parravano2003} to derive a map of the local FUV flux induced by the distribution of O/B stars on Fig.~\ref{fig:fdisk_facc_radec_map}, and examined any potential correlations between the calculated flux levels and the observed variations on $f_{disk}$ and $f_{acc}$. A direct comparison of these quantities is hampered by independent and competing factors such as age trends and age uncertainties in disk evolution or the mass dependence of disk timescales, and no conclusive evidence could be extracted from our bi-dimensional analysis. However, when restricting this exploration to K-type stars (which constitute the only sufficiently numerous, similar-mass subset in our sample to retain statistical representation after accounting for stellar age and spatial binning), $f_{disk}$ values larger than 0.5 tend to be associated with local FUV fluxes that are about 0.3~dex less intense than the typical estimate for spatial positions where the estimated $f_{disk}$ is smaller than $f_{evolved}$. This tentative distinction emerges in particular at ages $\gtrsim$1~Myr, although the derived typical values for high-$f_{disk}$ and low-$f_{disk}$ locations would still be consistent within the associated scatter.}

\section{Summary and Conclusions}\label{sec:conclusions}

In this work, we have examined the properties and evolution of disk accretion in young stars across the Lagoon Nebula region. Our analysis was anchored to a sample of 1012 member YSOs, distributed in mass between $\leq 0.2$\,$M_\odot$ and $\geq 5$\,$M_\odot$.
We employed homogeneous, simultaneous $g,r,i$ photometry, obtained with VST/OmegaCAM, to determine the fundamental stellar properties for individual objects ($A_V$, SpT, $M_\star$). We then combined the optical stellar colors with $u$-band and $H\alpha$ flux measurements (also obtained at the same time with OmegaCAM) to detect and measure excess emission linked to mass accretion onto the star.

By using near- to mid-IR archival photometry, we estimated a statistical disk fraction ${f}_{disk} \sim 0.34-0.37$ across our sample, which is consistent, within the uncertainties, with the derived fraction of YSOs with evidence of ongoing mass accretion, ${f}_{acc} \sim 0.38-0.41$. When broken down into spectral classes {and disk diagnostics}, statistically consistent values of $f_{disk}$ and $f_{acc}$ are measured {for sources with disk detection at mid-IR (3.6--8.0 $\mu$m) wavelengths}. {On the other hand, the estimated fraction of accretors appears to be systematically larger than the fraction of disks detectable in the near-IR (1.2--2.2 $\mu$m). This suggests ongoing inside-out disk clearing mechanisms that may be operating on quicker timescales around} higher-mass {(SpT $<$ K)} stars, although our surveyed population is estimated to be significantly incomplete at very low masses ($M_\star \lesssim 0.5$\,$M_\odot$). A close examination of spatial variations in the proportion of disk--bearing, disk--free, and accreting sources across the region suggests that dust depletion {may} occur more quickly for disks close to the core of NGC~6530, where the bulk of massive stars can be found. In the same area, the fraction of accreting stars appears to be similar to the fraction of disk--bearing stars, indicating that the quicker dust depletion does not immediately halt stellar gas accretion. Conversely, in the outer areas of the region, particularly along the southwestern star formation front, the fraction of disk--bearing stars tends to exceed the fraction of accreting stars and that of disk--free stars, suggesting a more regular disk evolution timeline. This differential evolutionary pattern for disk--bearing YSOs in the core vs. the outskirts of the region, in spite of their estimated similar ages, is also supported by the typical $\dot{M}_{acc}$ values measured in the two areas, higher by a factor of two along the southwestern rim than in the cluster center.

The distribution of mass accretion rates as a function of stellar mass exhibits an overall correlation with {at least one, and potentially} two, break points in the relationship: the first around $M_\star \sim {1.02}$\,$M_\odot$, and the second {tentative one} at $M_\star \geq {4.07}$\,$M_\odot$. As reported in earlier studies of accretion variability in YSOs, the variability amplitudes on $\dot{M}_{acc}$ detected here on timescales of a few weeks are statistically consistent with changes in the accretion rates measured {over timescales} of years for the same sources. This indicates that timescales of days to weeks typically dominate the dynamics of the targeted YSOs over baselines that are at least tens of times longer. The most erratic variables identified with {\em K2} tend to exhibit higher $\dot{M}_{acc}$ than the bulk of the population at a given mass; in addition, especially in the burster category, they tend to display smaller amounts of $\dot{M}_{acc}$ variability than the bulk of YSOs, which may indicate a more chaotic distribution of surface accretion shocks (leading to less modulation). The shocks themselves appear to be structured in density, leading to slightly asynchronous light curves in different filters and to potential differences in cycle duration as a function of wavelength. Indeed, a limited timing analysis of brightness phases from the $u$-band to the $i$-band, {compared} to the simultaneous {\em K2} light curves for accretion modulated stars, resulted in frequent lags being detected between the light curves in different filters. Measured delays are typically on the order of $\sim7\%$ of the optical light curve period extracted from the {\em K2} data, but can reach fractions as large as $>30\%$. 

Additional factors, such as stellar multiplicity, can impact the evolutionary picture for accretion disks examined here. Indeed, observational surveys conducted in other star formation environments have suggested that the presence of companions may lead to a rapid disk disappearance in the first few Myr of pre-main sequence evolution \citep[e.g.,][]{harris2012}, including dynamical disk disruptions that may trigger a quick depletion of the dust content in the inner disk \citep[e.g.,][]{francis2020}. To explore the connection between stellar multiplicity and inner disk dynamics, we have undertaken a survey of binary status for disk--bearing and disk--free young stars in the Lagoon Nebula that encompass a wide range of identified \textit{K2} variability behaviors. Results from that survey will be presented in a forthcoming paper (Venuti et al., in preparation).

\section*{acknowledgments}
We thank the anonymous referee for their thorough review that helped us strengthen the presentation of our results. This work was supported by the National Aeronautics and Space Administration (NASA) under grant No. 80NSSC21K0633 issued through the NNH20ZDA001N Astrophysics Data Analysis Program (ADAP). L.V. warmly acknowledges the Centre for the Subatomic Structure of Matter for their hospitality during her stays at the University of Adelaide as a Visitor in 2022--2023. This research made use of Lightkurve, a Python package for Kepler and TESS data analysis \citep{lightkurve}, and Photutils, an Astropy package for detection and photometry of astronomical sources \citep{larry_bradley_2022_6825092}. This publication makes use of data products from the Two Micron All Sky Survey, which is a joint project of the University of Massachusetts and the Infrared Processing and Analysis Center/California Institute of Technology, funded by NASA and the National Science Foundation. This publication also makes use of data products from the Wide-field Infrared Survey Explorer, which is a joint project of the University of California, Los Angeles, and the Jet Propulsion Laboratory/California Institute of Technology, funded by NASA. This work has also made use of data from the European Space Agency (ESA) mission
{\it Gaia} (\url{https://www.cosmos.esa.int/gaia}), processed by the {\it Gaia}
Data Processing and Analysis Consortium (DPAC,
\url{https://www.cosmos.esa.int/web/gaia/dpac/consortium}). Funding for the DPAC
has been provided by national institutions, in particular the institutions
participating in the {\it Gaia} Multilateral Agreement.

\vspace{5mm}
\facilities{VLT Survey Telescope (OmegaCAM), Kepler/K2, Spitzer (IRAC)}

\software{TOPCAT \citep{topcat}, SciPy \citep{scipy}, Photutils \citep{larry_bradley_2022_6825092}}

\bibliography{references}{}
\bibliographystyle{aasjournal}

\end{document}